\newif\ifcrc            

\crctrue

\def\url#1{{\ttfamily\def\/{/\discretionary{}{}{}}#1}}

\ifcrc
\documentclass[fleqn,twoside]{article}
\usepackage{espcrc2}

\usepackage{harvard}    
\usepackage{graphics}   %
\usepackage{epsf}       

\usepackage{amssymb}

\begin{document}

\title{MODEST-1:
\hbox{Integrating Stellar Evolution and Stellar Dynamics}}
\author{Piet Hut
\thanks{Corresponding author.\newline
{\it E-mail addresses: } piet@ias.edu (P. Hut), mshara@ amnh.org (M. Shara), sverre@ast.cam.ac.uk (S. Aarseth), rklessen@aip.de (R. Klessen), lombardi@vassar.edu (J. Lombardi), makino@astron.s.u-tokyo.ac.jp  (J. Makino), steve@kepler.physics.drexel.edu (S. McMillan), O.R.Pols@phys.uu.nl (O. Pols), teuben@astro.umd.edu (P. Teuben), webbink@astro.uiuc.edu (R. Webbink)}
\address{Institute for Advanced Study,
Einstein Drive, Princeton, NJ 08540, USA}
, Michael M. Shara
\address{Department of Astrophysics, American Museum of Natural
History, New York, NY 10024, USA}
, Sverre J. Aarseth
\address{Institute of Astronomy, University of Cambridge, Cambridge, UK}
, Ralf S. Klessen
\address{Astrophysikalisches Institut Potsdam,
An der Sternwarte 16, 
14471 Potsdam, Germany}
, James C. Lombardi Jr.
\address{Department of Physics and Astronomy,
Vassar College,
124 Raymond Avenue,
Poughkeepsie, NY 12604, USA}
, Junichiro Makino
\address{Department of Astronomy, University of Tokyo, 7-3-1 Hongo,
Bunkyo-ku, Tokyo 113-0033, Japan}
, Steve McMillan
\address{Department of Physics,
	Drexel University,
	Philadelphia, PA 19104, USA}
, Onno R. Pols
\address{Astronomical Institute, Utrecht University, P.O. Box 80000,
3508 TA Utrecht, The Netherlands}
, Peter J. Teuben
\address{Astronomy Department,
University of Maryland,
College Park, MD 20742, USA}
, Ronald F. Webbink
\address{Department of Astronomy, University of Illinois at Urbana-Champaign,
1002 W. Green St., Urbana, IL 61801, USA}}

%
%
\else
%
%

\documentclass{elsart}

\usepackage{harvard}    
\usepackage{graphics}   %
\usepackage{epsf}       


\usepackage{amssymb}

\begin{document}

\begin{frontmatter}
\title{MODEST-1:
\hbox{Integrating Stellar Evolution and Stellar Dynamics}}
\author{Piet Hut}
\address{Institute for Advanced Study,
Princeton, NJ 08540, USA}

\author{Michael M. Shara}
\address{Department of Astrophysics, American Museum of Natural
History, New York, NY 10024, USA}

\author{Sverre J. Aarseth}
\address{Institute of Astronomy, University of Cambridge, Cambridge, UK}

\author{Ralf S. Klessen}
\address{Astrophysikalisches Institut Potsdam,
An der Sternwarte 16, 
14471 Potsdam, Germany}


\author{James C. Lombardi Jr.}
\address{Department of Physics and Astronomy,
Vassar College,
124 Raymond Avenue,
Poughkeepsie, NY 12604, USA}

\author{Junichiro Makino}
\address{Department of Astronomy, University of Tokyo, 7-3-1 Hongo,
Bunkyo-ku, Tokyo 113-0033, Japan}

\author{Steve McMillan}
\address{Department of Physics,
	Drexel University,
	Philadelphia, PA 19104, USA}

\author{Onno R. Pols}
\address{Astronomical Institute, Utrecht University, P.O. Box 80000,
3508 TA Utrecht, The Netherlands}

\author{Peter J. Teuben}
\address{Astronomy Department,
University of Maryland,
College Park, MD 20742, USA}

\author{Ronald F. Webbink}
\address{Department of Astronomy, University of Illinois at Urbana-Champaign,
1002 W. Green St., Urbana, IL 61801, USA}




%
%
\fi
%
%

\begin{abstract}

We summarize the main results from MODEST-1, the first workshop on
MOdeling DEnse STellar systems.  Our goal is to go beyond traditional
population synthesis models, by introducing dynamical interactions
between single stars, binaries, and multiple systems.  The challenge
is to define and develop a software framework to enable us to combine
in one simulation existing computer codes in stellar evolution,
stellar dynamics, and stellar hydrodynamics.  With this objective,
the workshop brought together experts in these three fields, as well
as other interested astrophysicists and computer scientists.
We report here our main conclusions, questions and suggestions for
further steps toward integrating stellar evolution and stellar
(hydro)dynamics.
\vspace{1pc}
\end{abstract}

\ifcrc

\maketitle

\else

\end{frontmatter}
\vfill\eject
\tableofcontents
\vfill\eject

\fi



\section{Introduction}
\label{intro}

Population synthesis models have been used successfully in comparisons
with observations of the global properties of stars, star clusters,
and galaxies.
The simplest models are constructed from a weighted sum of individual
stellar evolution tracks, while more detailed models incorporate
some additional information about binary stellar evolution.

For some stellar environments such a synthesis approach is perfectly
adequate, and there the main challenge is to deal with the
considerable complexities of binary star evolution.  However, the
situation is very different for the class of {\it dense stellar systems},
defined as environments in which a typical star has a significant
chance to interact and possibly collide with another star during its
lifetime.  In such an environment stars of different ages can exchange
mass, disrupt each other or merge, and their merger products can get
involved in similar interactions; binary stars can encounter single
stars as well as other binaries, where one or more of the stars may
already be a merger product; and so on.  Examples of dense stellar
systems are star-forming regions and the dense cores of open and
globular clusters, as well as galactic nuclei.

It is clear that the possibilities are almost endless.  While population
synthesis based on single-star evolution can easily be exhaustive, and
synthesis based on a mixture of single stars and binaries can at least
aim to be reasonably complete, there is no way that one can anticipate
and tabulate all possible multiple-star interactions in dense stellar
systems.  Detailed attempts at population synthesis for such systems
by necessity have to be dynamical, taking into account the particular
ways that stars encounter one another in a given simulation.

During the last few years, several dynamical population synthesis
studies have appeared ({\it cf.} Portegies Zwart {\it et al.} 2001,
Hurley {\it et al.} 2001).  In these studies, the dynamics of a
dense stellar system is modeled through direct $N$-body integration,
while the stellar evolution is modeled through fitting formulae that
have been obtained from large numbers of individual stellar evolution
tracks.  Binary stellar evolution is modeled through the use of
semi-analytic and heuristic recipes (Hurley {\it et al.} 2002).

Astrophysically these results are novel and exciting, but their
reliability is not so easy to assess.  Validation is a core issue
here, requiring not only detailed internal checks but also comparison
between different codes run by different groups.  This question was
discussed at some length last year at IAU Symposium 208 in Tokyo,
resulting in the specification of a well defined set of initial
cluster and stellar parameters (Heggie 2002).  Given the fact that the
necessary codes are rather complex, requiring years of development,
so far few groups have been able to confront this new challenge.  This
stands in contrast to the first collaborative experiment (Heggie {\it
et al.}  1998), which was confined to stellar dynamics (without
stellar evolution), and attracted "entries" from about 10 groups.  We
hope that our new MODEST initiative will stimulate more groups to
engage also in the friendly competition of the second collaborative
experiment.

Further improvement to the more comprehensive simulations referred to
above will require the use of ``live'' stellar evolution models before
too long, in order to deal with the unusual types of new stars that
can be formed by mergers in dense stellar systems.  However, the
challenges of coupling existing stellar evolution codes and stellar
dynamics codes are quite daunting.  The first workshop specifically
organized to address these challenges was held during July 17-21, 2002
at the American Museum of Natural History in New York City.  The
workshop brought together a group of experts in stellar evolution,
stellar dynamics, stellar hydrodynamics and other fields of
astrophysics, as well as computer scientists.

Originally, the workshop was announced to a small group of people who
were known to work on the interface of dynamics and evolution, under
the title ``Integrating Stellar Evolution and Stellar Dynamics''.  We
originally expected to see a handful of participants for an informal
round-table discussion.  The fact that instead 34 attendants convened
is a clear sign of the timeliness of the meeting, and the desirability
to form a concerted effort to bridge the gap between the stellar
evolution and dynamics communities.

This paper offers a summary of the week-long series of discussions
held during the workshop, distilled by the organizers (Piet Hut and
Mike Shara) and eight of the participants representing a cross section
of expertise available during the meeting.  In addition, we have
created a web site\footnote{\tt http://www.manybody.org/modest.html}
where the name `modest' reflects our renaming of the meeting during
the last day to MODEST-1, the first workshop on MOdeling DEnse STellar
systems.  We plan to hold biannual follow-up meetings, MODEST-2 in
Amsterdam in December 2002, and MODEST-3 in Australia in July 2003.
In addition, we have started an email list to facilitate ongoing
discussions about technical details of dynamical population synthesis
simulations.  Further information can be found on our web site.

As a summary of our workshop, this paper contains the input of all of
the participants, which are listed below under the acknowledgments.
While many of the authors have contributed to various sections, each
section has one or two main authors, as follows.  \S1 and \S4 were
written by Piet Hut, \S2 by Michael Shara, \S3 and \S6 by Piet Hut and
Jun Makino, \S5 by Onno Pols and Ronald Webbink, \S7 and \S8 by James
Lombardi, \S9 by Sverre Aarseth and Ralf Klessen, \S10 by Steve
McMillan and Peter Teuben, and \S11 by Steve McMillan.

In order to make the discussion concrete we have provided specific
code fragments in \S6 and \S8 below.  We see this paper as the start
of a discussion that will ultimately result in the definition of clear
standards for interfaces between stellar dynamics, evolution, and
hydrodynamics.  However, the current fragments are for illustration
only, and are {\em not} necessarily intended to become part of any
future standard.

\section{Predictions}
\label{observations}

The successful marriage of $N$-body simulations with increasingly
sophisticated stellar evolution codes of all flavors will yield
progeny whose genetic characteristics should be designed now, to avoid
petabytes of untestable output.

Essential ingredients of science are predictions and testability.  Of
course, we all look forward to detailed models of star clusters with
self-consistent stellar evolution spanning aeons of time. But we want
to emphasize how critical it is to generate those models with enough
genetic markers to allow observers to tell us if our models have
anything at all to do with physical reality.

A poster child for this kind of approach is the important paper by di
Stefano and Rappaport (1994), where directly testable
predictions of the cataclysmic binaries in a few selected globular
clusters were made. Such predictions are dangerous for the egos of
theorists (it's not fun when observers find many orders of magnitude
more or less than what you predicted) but it's essential to the health
of our science.

Modelers of star clusters are confronted with datasets rich in genetic
markers from HST , Chandra and other observatories. Detailed sequences
of blue stragglers, white dwarfs, X-ray binaries, millisecond pulsars
and ``missing'' red giants are now available for significant numbers of
globular clusters. While these often represent less than ten percent
of the cluster (both in terms of numbers and in terms of mass), they
must be reproduced in the correct numbers and positions in clusters if
we are to have any confidence in the coming generations of MODEST
models.

A slightly more subtle, but no less important set of predictions that
should be made by combined $N$-body and stellar evolution codes
concerns the lineages of tracer stars. It is just as informative to
know how each blue straggler in a cluster got that way as it is to
know how many blue stragglers are predicted in a cluster. The ``synthetic
history'' of each star should not be taken literally because of the
chaotic nature of the individual particle trajectories.  However,
the cumulative, statistical histories of entire classes of stars are
important because these make testable predictions.

A concrete example comes from recent simulations of
Shara and Hurley (2002) of M67-like star clusters. The
life cycle of every white dwarf binary in every simulation was
followed in detail, focusing on the systems that eventually
merge. The key result is that the white dwarf merger rate is enhanced,
relative to the field, by over an order of magnitude. The life story
of any particular binary white dwarf in this simulation isn't
important. However, the history of the entire class of objects is very
important: it directly predicts that SNIa may be preferentially
produced in star clusters. This is observable and hence testable.

In summary, theorists should consider providing not just the numbers,
lifetimes, luminosities, colors and spatial distributions of every
class of ``tracer'' star in a cluster.  These will be indispensable in
directly matching observed clusters to simulated clusters.  But deeper
insights into the evolution of star clusters can be gained by
retaining statistical information about the histories of stellar
populations from the $N$-body with stellar evolution simulations.

\section{A MODEST Approach}
\label{approach}

\subsection{Divide and Conquer}

Conceptually, it would be easiest to start from scratch in order to
model the gravitational, hydrodynamic, and internal interactions
between stars.  In such an approach, one could choose a particular
computer language and style of programming, define the appropriate
data structures and abstraction barriers, and write the various parts
of the program accordingly.  And indeed, such a project might be
feasible, but would probably take a team of people years to accomplish.
For the near future it makes more sense to work with existing computer
codes that already can handle the dynamics or evolution or hydrodynamics
that are needed to model dense stellar systems.

For one thing, many of these three types of programs already
incorporates tens to hundreds of person-years of collective
experience, and it will be far from easy to codify and reproduce that
expertise, much of which has never been formalized, and some of which
may never even have been commented properly.  For another, we
literally have no experience at this point in setting up large-scale
attempts at integrating these various physical aspects in simulations
of dense stellar systems.  Given this situation, it would seem most
prudent to start experimenting with existing codes, matching them with
toy models first, and then with each other, in order to gain some
initial experience concerning their collective behavior.

Our MODEST acronym lends itself very conveniently to express this aspect of
our philosophy: our approach is one of MODifying Existing STellar codes.
We hope this reading will avoid the false impression that either we or
our projects could possibly be considered modest.

The main price to pay for MODESTy is that we have to find ways to
connect bits and pieces of code that are written in different computer
languages, and in different styles, ranging from the use of Fortran
before the invention of subroutines to the use of highly structured
object-oriented languages such as C++ or even lisp dialects, and
possibly scripting languages such as Perl, Python or Ruby.

A bonus of the MODEST approach is that working with black boxes as
components allows a swapping of those black boxes, which will make
validation of the final results much easier.  If we can easily change
the use of one stellar evolution code for another, for example, we can
quickly get an impression of the relative accuracy of those codes (to
the extent that they are truly independent).  Such a divide-and-conquer
approach is crucial in proving correctness of the outcome of highly
complex large-scale simulations.

To sum up, the challenge is to construct a software framework that
allows us to model a wide variety of astrophysical situations,
using existing programs that encapsulate specialized astrophysical
expertise.  Where necessary, we will write wrappers, drivers, and
other modules that will communicate and translate information between
the already existing programs.  What is needed first is to define a
convenient and well-specified set of interfaces that allow us to mix
and match the various unrelated programs, written in different
languages and in different styles, in such a way that they can appear
as black boxes to each other and to one or more driver programs.

\subsection{Specification of Interfaces}

A central task in setting up a software framework for any type of
large-scale simulation is the specification of interfaces between
different computational modules.  On the one hand, we must be careful
not to force any particular organization on the variables that are
private to each module.  On the other hand, we should maintain
consistency across an interface.

In general, for each interface there should be an agreement about the
particular names and types of a minimal set of variables that will be
passed through the interface.  This does not mean that the modules
themselves will be forced to use those externally constrained names
and types; it is straightforward to provide extra levels of data
abstraction, for example by writing wrappers around existing modules
that translate the information from the relevant variable within the
module to the names and types specified in the interface.

It also does not mean that interface specifications will be put in
stone.  On the contrary, an essential aspect of good interface design
is to leave open the possibility of significant future extensions of
what will be passed through an interface, perhaps totally unforeseen at
present.  The only requirement will be compatibility with older
specifications of the interface.

In the concrete case of simulations of dense stellar systems we have
three broad classes of existing programs that already model aspects of
astrophysical phenomena.  These are stellar dynamics, stellar
evolution and stellar hydrodynamics.  In the future, we may want to
write a special driver/scheduler/manager program, but in existing
stellar dynamics programs, more than 90\% of the lines of code are
already dedicated to such orchestration details.  Therefore,
initially at least it will be simplest to consider the evolution and
hydrodynamics programs as black boxes that are invoked by the dynamics
program when needed.  Later implementations may grant a more active
role to the evolution and hydrodynamics programs, if that would reduce
complexity and dependencies.

Given that current codes are written in totally different styles and
in different languages, our first task is to specify interfaces and to
develop wrappers around existing programs that are compliant with
those interfaces.  Since we all have different backgrounds, we can
help reach this goal in different ways, according to what we enjoy
doing and what we're already good at.  It would be counterproductive
to require a specialist in stellar evolution to suddenly learn new
computational science tools he or she is not comfortable with; similarly
it would be counterproductive to require a stellar dynamicist to
become familiar with the inner details of how a stellar evolution code
is set up.  If someone has programmed in Fortran for thirty years,
there is absolutely no reason to require this person to learn and use
other languages (although it might be fun).  There is even no need for
that person to do any work on writing the wrapper around his or her
program for the interface; {\em the minimum collaboration needed is a clear
specification of which variables in his or her program correspond to
those specified for the interface.}  Providing those in a Fortran
common block, say, would be fine if that is the style this person is
used to program in.

\section{Stellar Dynamics}
\label{dynamics}

The earliest published $N$-body simulations are the 10-body runs by von
Hoerner \cite{vH60}.
By the early seventies, larger systems could be modeled, up to $N=500$.
Key ingredients in making it possible to integrate these larger
systems were the use of individual time
steps \cite{A63}, as well as special treatments of binaries
through various ways of analytical and other forms of regularization
\cite{A85}.  
The two leading families of $N$-body codes tailored to simulations of
dense stellar systems are {\tt NBODYx}\footnote{NBODY4 is optimized
for use on the GRAPE special-purpose hardware; other members include
NBODY6 for general-purpose single processor computers, and NBODY6++
for parallel computers} (Aarseth 2002, Spurzem and Baumgardt 2002),
and the {\tt kira} integrator distributed with the
Starlab\footnote{\tt http://www.manybody.org/starlab.html} software suite
(Portegies Zwart {\it et al.} 2001; Hut 2002).
For a general treatment of dense stellar systems, and especially of
rich star clusters, see Heggie \& Hut (2002).

Hardware improvements were important as well, in reaching the goal of
simulating whole star clusters.  The GRAPE project of constructing
special-purpose computers, initiated at Tokyo University in 1989, has
led to the installment of dozens of such computers world wide.  An
example of calculations made possible by the GRAPE was the first
demonstration of the occurrence of core oscillations in direct
$N$-body systems by Makino (1996), using the GRAPE-4 to perform a
32,000-body calculation.  The acronym ``GRAPE'' stands for GRAvity
PipE; more information can be found on the GRAPE web
site,\footnote{\tt http://www.astrogrape.org} in the book by Makino \& Taiji
(1998), and the review articles by Hut \& Makino (1999) and Makino (2002).

\subsection{The Physical Role of Stellar Dynamics}

Stellar dynamics is perfectly adequate in modeling the motions of
stars as point masses moving under the influence of gravity, even in
dense stellar systems, unless individual stars approach each other to
within a few stellar radii.  When that happens, the internal structure
of the stars has to be taken into account, and we have to switch to a
hydrodynamics module to follow the encounter, which may lead to mass
transfer and even to the merging of two or more stars.  After the dust
has settled, we then have to update the stellar evolution models for
the stars involved, and in case of mergers we will have to construct
new models from scratch, often with highly unusual chemical
compositions.  All of this has to happen automatically, which means
that the individual modules have to be robust, and that the interfaces
should be well-defined.

\begin{figure}
\begin{center}
\ifcrc
\epsfxsize = 3.0in
\else
\epsfxsize = 4.5in
\fi
\epsffile{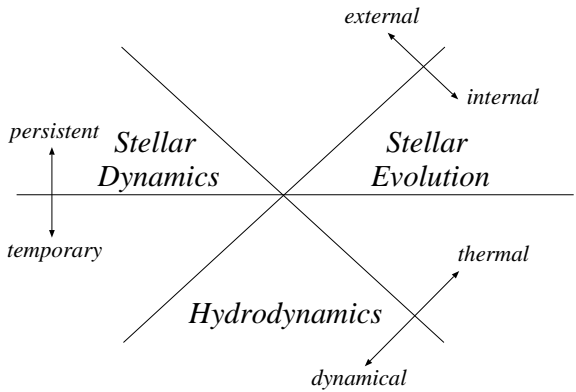}
\caption
{Three aspects of simulations of dense stellar systems, and three ways
to classify them into two categories.}
\label{fig:fig1}
\end{center}
\end{figure}

The three types of physics involved in stellar interactions are
sketched in Fig. 1.  Each type plays a unique role in terms of type of
degrees of freedom, time scale, and duration.  For example, stellar
dynamics is concerned with external degrees of freedom, on a dynamical
time scale, and for the duration of the whole history of the star system.

To start with the first distinction: the hydrodynamics and evolution
codes are only concerned with the internal degrees of freedom of the
stars, whereas the stellar dynamics module orchestrates the evolution
of the external degrees of freedom: positions and velocities and
higher derivatives thereof.  The dynamics needs to know the masses
and radii of the stars,\footnote{This simple distinction may become
blurred when more complex dynamical processes, such as tidal
interactions and possibly tidal capture, are considered.}
the masses to compute gravitational forces and the radii to warn for
possible collisions, but it only actively updates the positions (and
velocities, accelerations, jerks, etc.) of the system.

A second distinction is given by the time scales on which the different
processes evolve the stars.  Stellar dynamics and hydrodynamics both
use explicit integration schemes in order to follow the stars on a
dynamical time scale.  Stellar evolution codes, in contrast, use
implicit integration schemes to follow the changes in internal
structure of a star on thermal and nuclear time scales.  The physical
reason is that dynamical equilibrium can be assumed to be accurately
preserved during almost all stages of stellar evolution.  In contrast,
it is exactly the deviation from dynamical equilibrium that drives the
hydrodynamical phenomena.  The situation is intermediate in the case
of stellar dynamics: a Fokker-Planck code, for example, follows a
star system on a ``thermal'' (two-body relaxation) time scale, but direct
$N$-body codes follow all stars on a dynamical time scale, which is
necessary to accurately model phenomena involving binaries and
multiple star systems.

The third distinction concerns the duration of the relevance of each
physical process.  Each star in the system will always be represented
as a point mass in the stellar dynamics part of the code, and as a
star with internal structure in the stellar evolution part of the
code.  While these two representations persist throughout the full
history of a simulation, the third type of representation, offered by
a hydro code, is temporary.  Only during a close encounter do
hydrodynamical models for a few stars spring to life, and they are
again discarded after they have done their duty, after a period
comparable to a few crossing times of the system (a day or so for
normal stars, a year at most in the case of giant stars).

Note that this description only applies after the stars have been formed.
During the earlier stages, when a star system is born through the
collapse of molecular clouds, hydrodynamics also plays a more global role.
Like stellar dynamics, the hydrodynamics describes the external
degrees of freedom of the gas clouds, and it is a persistent element
in the computer code for the simulation, as long as gas remains
present in the system.  See \S9 for more details.

\subsection{The Computational Role of Stellar Dynamics}

When we compare the complexity of the three physical processes, it is
clear that stellar dynamics is by far the simplest, conceptually.  The
only computational task is the integration of Newton's classical
gravitational equations of motion.  What could be simpler?  In
comparison, the dynamical fluid equations of hydrodynamics are far
more subtle, largely because they are partial differential equations
rather than ordinary differential equations.  The possible occurrence
of shock waves and turbulence has no analogy in the simple world of
stellar dynamics.  And the intricacies of stellar evolution are even
more subtle, with the interplay between radiative transfer, nuclear
energy generation, convection, the largely still unknown roles of
rotation and magnetic fields, and so on.

Given this situation, why are state-of-the-art stellar dynamics codes
so complex, and why are they still being improved, after forty years
of collective experience in writing them?  They answer lies in the
fact that what we call a stellar dynamics code is in fact mostly a
complex scheduling manager where almost all the logic is used to make
sure that the integrations retain accuracy.  In the thousands of lines
of computer code in a modern stellar dynamics program, only a few
hundred lines contain Newton's force calculation and the integration
thereof.  All the rest of the code involves special forms of treatment
for each star.

For instance, unlike almost all text book examples of the integration
of differential equations, stars in $N$-body systems are integrated
with individual time steps.  In addition, close encounters between
stars are treated in special ways, by constructing local coordinate
systems to represent their positions in order to avoid round-off
errors.  Not only does the proper creation and destruction of these
coordinate patches require quite a bit of intelligence in a dynamics
code, the real fun starts when two or more such coordinate patches
meet, and have to merge or split.  And on top of all that, specific
code is often written to avoid the numerical singularities involved in 
close encounters of stars, for example by adopting special treatments
of unperturbed motion, or mapping the three-dimensional Kepler motion
onto that of a four-dimensional harmonic oscillator through the
Kustaanheimo-Stiefel transformation ({\it cf.} Aarseth 2002).

For all these reasons, the structure of a computer program that can
model stellar dynamics, stellar evolution, and hydrodynamics is not
well described by the schematic diagram in Fig. 1, that focuses
only on physical processes.  Instead, we can discriminate between
three different aspects of a typical stellar dynamics program for
dense stellar systems.  The most straightforward part of the program
governs the integration of the global objects in the system.  These
objects can be single stars, isolated binaries, triples or higher
multiples, as well as temporarily interacting groups of stars.  Each
non-trivial object (anything that is not a single star) has additional
internal gravitational degrees of freedom.  For example, an isolated
binary might be represented through an analytic expression in the form
of a Kepler orbit, which can be used to predict the position of the
stars when they are needed, during a relatively close encounter.  And
the dynamics of an interacting group of stars will be computed using
its own local coordinate system, possibly using regularization methods.

Besides this division of labor between global and local gravitational
interactions, each stellar dynamics code contains a third segment in
the form of a piece of code that takes care of the overall scheduling
of all events that occur.  This scheduler acts as a system clock that
tells each particle when it has to move (remember that different
particles have different time steps), and in addition it issues the
orders for the creation and destruction of local coordinate patches,
as well as their merging and splitting.  Therefore, from a computational
point of view, the neat division into three different physical processes
translates into the five different computational processes sketched in
Fig. 2.

\begin{figure}
\begin{center}
\epsffile{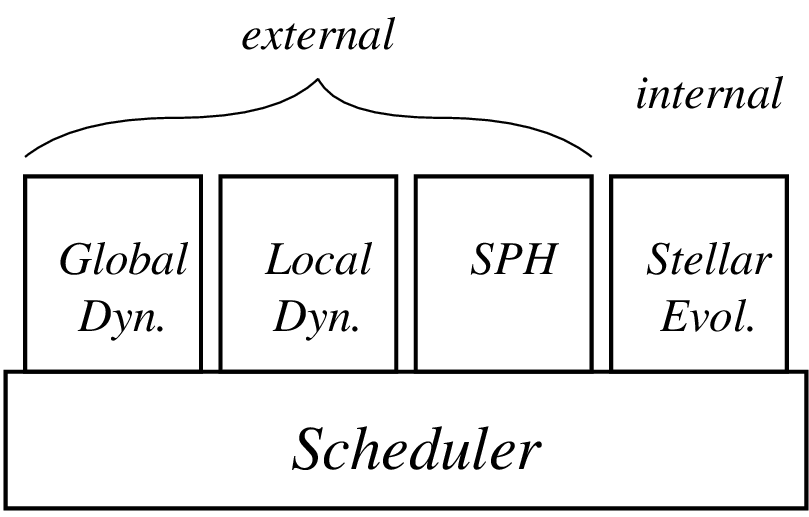}
\caption
{The stellar dynamics part of a combined simulation code contains
three different parts: one modeling the global dynamics, one for the
local dynamics, and an overall scheduler.  It might be most natural
to let the scheduler do the synchronization for the hydrodynamics and
stellar evolution modules as well.}
\label{fig:fig2}
\end{center}
\end{figure}

The same logic that is in place in current stellar dynamics codes
already effectively contains an interface between the local and global
part of the gravitational calculations, as well as a mechanism for the
partly asynchronous evolution of the various components.  It would be
natural to use these features, in our philosophy of trying to make
only minor modifications to existing stellar codes, as stressed in
\S3.  Such a strategy would lead to the following requirements, at
least for initial progress in realizing the physics of Fig. 1:

\begin{enumerate}

\item
      make a clear and clean separation between the local and
      global gravitational components of current stellar dynamics
      codes.

\item
      make a clear and clean separation of the synchronization
      part of such a code from the rest of the dynamics.

\item
      specify interfaces between these three parts of a dynamics code,
      in order to allow a homogeneous treatment between those
      interfaces (currently internal in dynamics codes) and interfaces
      with the external modules that govern hydrodynamics and evolution.

\item
      construct interfaces between the scheduler and the hydrodynamics
      and stellar evolution modules along similar lines as was done
      for stellar dynamics.

\item
      finally define the interface between stellar dynamics and
      hydrodynamics, as well as between hydrodynamics and stellar
      evolution, in such a way that the internal stellar properties
      can be modeled with a similar predictor-corrector structure as
      is currently done for the external variables in stellar dynamics codes.

\end{enumerate}

To clarify point 4): for the foreseeable future, it is probably most
efficient to represent a hydrodynamics module for a star through a
Smooth Particle Hydrodynamics (SPH) module, a form of an $N$-body code
where each particle is given an entropy in addition to a mass and
position and velocity.  Computationally, both hydrodynamical and stellar
dynamical degrees of freedom are then modeled as external degrees of
freedom, while the stellar dynamics is still modeled as a black box
with internal degrees of freedom.  When two or more stars come close,
their gravitational point mass external information, together with
their stellar evolution internal information, are used together to
construct a temporary hydrodynamic representation.  After the encounter,
the hydrodynamic information is translated back left and right into the stellar
dynamics and evolution modules.  With the hydrodynamic module as a
go-between, there may be no need for the stellar dynamics and stellar
evolution modules ever to talk to each other directly.

To clarify point 5): note that an $N$-body code is not really a
code that follows $N$ point masses.  Rather, its internal representation
deals with $N$ orbit segments.  Each star has a position, velocity and
higher derivatives that have last been calculated at a give time.
Based on that information, the future orbit of the particle can be
predicted up to a particular later time, a type of `latest sales date'
for which the accuracy is guaranteed to stay within the required bounds.
As soon as the system time exceeds this later time, that particular star
will be updated, so that its `latest sales date' again is pushed into
the future, beyond the current time.  Until that new time is reached,
all other particles can once again rely on the newly computed orbit
segment to provide information about the given star when needed,
at times other than the time at which this star was updated.

This elaborate mechanism that makes it possible to advance stars at
individual time steps can be extended to the treatment of hydrodynamics
and stellar evolution as well.  What is needed in this case is a type
of interface that can ask the stellar evolution module, for example,
to provide an estimate of its near-future behavior, and a `latest
sales date' until which this information can be considered to be
accurate.  For a user at the stellar dynamics side of the interface,
it is irrelevant whether such a prediction is a true prediction, or
simply a reading of an entry in a table, or an actual calculation by
evolving a stellar evolution model for some duration into the future.
We will come back to these issues in \S6.

\section{Stellar Evolution}
\label{evolution}

\subsection{Background and motivation}

We require a code that models the evolution of any star, either
single or binary, from an arbitrary initial condition up to the end of its
nuclear and thermal evolution.  Such a code will have a wide range of
applications, but the main application we consider here is for modeling
dense stellar systems such as globular clusters, galactic nuclei or
starbursts, where many stars (of order 10$^5$ or more) interact with one
another and would have to be modeled simultaneously.  The requirements of
such a code are therefore: (1) it should be able to run autonomously and
without outside interference given a sufficient set of initial conditions
for the star(s); (2) it is robust and gives a -- hopefully meaningful --
result under any conceivable circumstance; (3) it is sufficiently fast that
an entire simulation of 10$^5$ stars or more takes a reasonable amount of
time (days at most); and (4) it should be able to interact with its
surroundings at any time, i.e. yield information about its current status
and also receive information that can modify its status.  These
requirements are by no means trivial!  At present no full-scale stellar
evolution code exists that satisfies all of these requirements, especially
points (1) to (3).  Every stellar evolution code of which we are aware is
prone to break down and needs to be nursed at some point between the
pre-main sequence and the white dwarf stage under the vast majority of
circumstances, and is certain to break down irretrievably under many
circumstances!  Besides the codes are still too slow, taking of the order
of a few minutes at least per star on the fastest processors available
today, so that a full simulation would take months.

Nevertheless, it should be pointed out that at the lowest level, at least
two codes that satisfy all four requirements are already in existence.
These are not full-scale evolution codes but rather parametrize stellar
evolution, using detailed evolution models as a basis wherever possible, and
making educated guesses otherwise.  These codes (i.e. \textsc{bse}, developed
by Tout et al. [1997] and Hurley, Tout \& Pols [2002]; and \textsc{seba},
developed by Portegies Zwart \& Verbunt [1996]) perform well and
have been successfully integrated into $N$-body codes (Hurley et al. 2001; 
Portegies Zwart et al. 2001).\footnote{in addition many other such
recipe-based binary evolution codes have been developed during the last
decade by various groups, but so far have not been used in conjunction with
$N$-body codes.}  They appear to give reasonable results under many
circumstances.  However there are circumstances, the most common of which --
in a dense stellar environment -- are probably the occurrence of mergers and
collisions, where the result of this approach probably has very little to do
with reality (cf. point 2).  It is especially with these collision products
in mind, as well as the fact that with 100,000 stars interacting something
unexpected and unparametrized is almost certain to happen, that we would
like to improve on these codes and make them more generally applicable.

It should be noted that a full-scale stellar evolution code, named
TYCHO, is freely available online\footnote{\tt
http://chandra.as.arizona.edu/$\sim$dave/tycho-intro.html}, courtesy of
D.\ Arnett.  TYCHO is an open-source, community code written in
Fortran.  For more details on the code, see \citeasnoun{you01} and
references therein.

In what follows we will discuss what has been done so far, what needs to be
improved, and how this can best be achieved.  We first discuss the situation
for single stars and then the more complex situation for binary stars.

\subsection{Single stars}

The theory of single-star evolution is rather well-developed, although
major uncertainties remain.  In particular convection can only be modeled
in a very crude way, while other internal mixing processes, e.g. induced by
rotation, have only begun to be explored.  The possible effect of internal
magnetic fields has hardly been studied at all.  Furthermore, mass loss is a
major uncertainty, especially for very massive post-MS stars, Wolf-Rayet
stars and AGB stars.  Nevertheless we are confident that single stars can
be modeled in a satisfactory way, and of all the uncertainties only
mass loss directly affects the dynamical evolution of a star cluster.

Single stars as they occur in dense stellar systems can be divided into
\emph{primordial stars}, which should evolve no different than single
field stars, and \emph{merged stars}, the products of collisions or
mergers, which may evolve quite differently from primordial stars.

\subsubsection{Current status and shortcomings}

In both codes mentioned above, \textsc{bse} and \textsc{seba}, single-star
evolution is modeled using a set of analytic formulae that have been
fitted to detailed stellar evolution tracks.  The \textsc{bse} code uses
the formulae constructed by Hurley, Pols and Tout (2000) that give several
global stellar quantities, such as luminosity $L$, radius $R$, mass $M$ and
core mass $M_c$, as a function of initial mass $M_0$, metallicity $Z$ and
age $t$.  Some of the formulae are rather ad hoc fits that reproduce the
shape of an evolutionary track in certain phases like the main sequence,
while others represent (in a simple way) actual physics underlying the
evolution, such as the core-mass luminosity relation that drives the
evolution of low-mass giants and AGB stars.  The fits also allow very fast
evaluation of certain important evolutionary timescales.  Mass loss is not
included in the fits but parametrized separately, so that different mass
loss prescriptions can be used in conjunction with the formulae.  The
formulae also provide other global quantities like the moment of inertia
and the depth of the convective envelope, so that the rotational evolution
can be modeled (if assumed rigid) as well as magnetic braking.

Although this approach has been applied successfully both in ordinary
(binary) population synthesis and in dynamical studies, it has several
shortcomings.  First and foremost, since the formulae have been fitted to
standard stellar models, they can only be expected to represent the
evolution of primordial stars.  Merged stars, on the other hand, are
expected to have rather different internal structure (i.e. composition
profiles) and so are probably not very well represented by the formulae.
As recent hydrodynamical studies have shown (see \S7), the
structure of collision products is neither homogeneous nor resembles that
of a primordial star with the same total mass.  Furthermore, the collision
products rotate very rapidly and are initially strongly out of thermal
equilibrium.  At present, however, the formulae are being used to represent
merger products and ordinary stars alike.

Also very massive primordial stars, whose evolution is determined to a
large extent by mass loss, are represented rather poorly because the
formulae are based on constant-mass models even though mass loss is taken
care of when applying the formulae.  Although such very massive stars only
form a tiny fraction of the initial population of a globular cluster, their
evolution and mass loss (which is poorly constrained in the first place) is
crucial for the early dynamical evolution of a cluster.  On the other hand,
for the other much more common types of star with very strong mass loss,
low- and intermediate-mass AGB stars, a formulaic approach is arguably the
\emph{best} way of representing their evolution.  Their evolution is driven
by a core-mass luminosity relation, which itself is not or only weakly
dependent on mass loss.  Mass loss is however crucial in determining the
lifetime of the AGB phase.

Another shortcoming, that becomes serious when we start modeling collision
and merger products in any detail, is that the formulae give no information
on the internal composition and entropy profiles.  Although it is possible
-- and useful for some purposes -- to represent and follow surface
compositions in a formulaic approach, deriving fitting formulae for entire
composition and entropy profiles is an extremely daunting task, and given
our experience with fitting even simple quantities like radius in a
satisfactory manner, a task that no one can realistically be expected to
carry out.  For the same reason, it is unlikely that the formulae will be
updated or replaced when a newer generation or extended set of stellar
evolution models becomes available.

\subsubsection{The best way forward}

Given these shortcomings, a different approach will need to be taken in
the future.  Here we must make a distinction between primordial stars
and merged stars.

Primordial stars, as argued above, all evolve alike for a given mass and
metallicity, if we neglect for the moment the possible effect of
(differential) rotation on the internal mixing processes.  Therefore the
most feasible approach, given the problems with speed and robustness of
current evolution codes, is to interpolate in a library of stellar models.
Such a library only needs two dimensions, $M$ and $Z$ (with a time-sequence
for each entry), and so is of manageable size.  For massive stars, perhaps
several libraries should be computed/compiled for different mass-loss
prescriptions.  On the other hand, for stars with a clear core-envelope
structure that follow a core-mass luminosity relation, i.e. AGB stars and
low-mass giants, parametrizing the evolution with analytic formulae
probably remains superior to table interpolation.  For these stars the
envelopes are homogeneous and (nearly) isentropic so it is sufficient to
follow the surface composition and entropy.  Hence a combination of
table interpolation and analytic formulae seems the best approach for the
near future.  

It should be noted that interpolation between stellar models is a
non-trivial task!  It is of the utmost importance that interpolation is
done between models in corresponding stages of evolution.  Hence these
evolution stages, at any rate the main critical turning points (e.g.
terminal-age main sequence, base of the giant branch, etc.), should be
identified on each track.  Furthermore internal composition and entropy
profiles need to be interpolated.  Constructing an interpolation routine
that can do all this automatically will be a difficult task, but the
advantage is that once it is available, it can handle any library so that
model libraries can be exchanged or updated at will.  An alternative
approach that circumvents the difficulties of interpolation is to use
discrete models from the library to represent a range of stellar masses.
For this to work the library has to be sufficiently densely spaced in
mass (and metallicity), i.e. masses not differing by more than a few
per cent.  This may be sufficient for modeling the dynamics, but if we
want to compare e.g. a color-magnitude diagram with observations we
may still wish to interpolate in order to prevent a discrete appearance.

As for merger products, these have been shown to have internal structures
quite unlike primordial stars.  The resulting structures from hydro
simulations rotate rapidly and are strongly out of thermal equilibrium, and
in order to relax they need to shed a large amount of angular momentum.  It
is not clear from the hydro calculations how this is achieved, nor can
stellar evolution codes answer this question, and this transition is likely
to remain a grey area for quite some time.  In any case it has to be
supposed that somehow the merged star manages to get rid of its excess
angular momentum, perhaps by shedding a small amount of mass in a disk.

It is also conceivable, but has yet to be verified by detailed
calculations, that the strong differential rotation leads to
additional mixing that significantly changes the chemical profile by
the time the star has relaxed to thermal equilibrium.  However, it
seems unlikely that the merger products will be completely
homogenized (\S7).  If the latter were the case, it would be conceivable to
construct an extended library of stellar models, with an additional
dimension namely the helium content, so that all of stellar evolution
could be done by table interpolation.  However, with arbitrary initial
composition profiles this clearly becomes impossible.

It seems therefore necessary to be able to do on-demand stellar evolution
calculations, for arbitrary initial entropy and composition profiles, during
a cluster simulation.  As discussed in Section 1, currently available codes
do not satisfy the demands that make integration into an $N$-body code
feasible.  Speed is one problem, although this will become less and less
important as processors get faster.  Nevertheless some effort will have to
go into making existing codes as fast as possible, by simplifying much of
the input physics like the equation of state, employing a minimal nuclear
network, and taking as few zones as is necessary to still achieve reasonable
accuracy.  In this way it should be possible, with current processors, to
evolve a star from the zero-age main sequence to the start of double-shell
burning in under one minute.  If only merger products are calculated this
way, this may not slow down a full $N$-body simulation too drastically,
particularly if many stars are computed in parallel on separate processors.

A much more daunting problem is robustness.  Although some codes can now
evolve unaided through the helium core flash and through many thermal pulses
along the AGB, this cannot be expected to be reliable under all
circumstances (in particular the unusual circumstances that we are
interested in).  Nor is it desirable, because each thermal pulse cycle takes
as much computing time as the entire evolution up to the first pulse.  So
even if codes can be made faster and more robust, we may still want to
parametrize the AGB phase, and perhaps skip over the He core flash by using
a set of pre-calculated zero-age horizontal branch models.

\subsection{Binary stars}

Binary star evolution presents a number of major problems of long standing
that have yet to be satisfactorily resolved (see, for example, Shore, Livio \&
van den Heuvel 1994).  A comprehensive description of how these problems are
dealt with (or circumvented) in one particular recipe-based binary evolution
code is given by Hurley, Pols \& Tout (2002).  We enumerate here 
only some of the more prominent ones.

Contact binaries, which are
characterized by large-scale energy exchange between components in their
common envelopes, account for nearly 1\% of solar-type stars in the galactic
disk (Rucinski 1998).  No established model exists for the physics of that
energy
exchange, even though in extreme cases it must account for as much as 99\% of
the energy radiated by the less massive star.  Evolutionary models of contact
binaries, using heuristic models of energy exchange, predict long phases of
semi-detached evolution which are not observed, or else demand such rapid
angular momentum loss in order to suppress the semi-detached state that they
cannot account for the abundance of contact binaries (see, for example, the
review by Eggleton 1996).  Not even heuristic
models exist for early-type contact configurations, which nevertheless arise
with startling frequency in models of more massive binary evolution (Nelson \&
Eggleton 2001). 

Evolutionary
analyses of individual Algol-type binaries (longer-period semi-detached
binaries with low-mass subgiant donors) almost invariably demand that they
have lost a significant fraction of their initial orbital angular momenta (and
possibly also significant fractions of the mass being transferred between
components) to have arrived at their present evolutionary state (e.g.,
Giuricin, Mardirossian, \& Mezzetti 1983; Eggleton \& Kiseleva-Eggleton 2002). 
It is widely
believed that magnetic stellar winds are responsible for these angular
momentum losses, and also those which drive unevolved binaries into contact
and cataclysmic binaries into mass transfer; yet the magnetic stellar wind
braking rates adopted in evolutionary calculations are almost invariably gross
extrapolations from the much weaker rates deduced empirically from rotation
velocities of main sequence stars in young star clusters.

The affinity of
other close, but not yet mass-transferring, binaries (the RS CVn binaries) for
nearly equal or slightly reversed mass ratios, strongly suggests that normal
stellar wind mass loss rates can be amplified significantly as stars approach
their Roche lobes (Popper \& Ulrich 1977; Tout \& Eggleton 1988), but no
physical model exists to quantify this amplification.

Field cataclysmic binaries (as also low-mass X-ray
binaries
and close double white dwarfs) are clearly products of dissipative evolution
within a more slowly rotating common envelope (Paczy\'{n}ski
1976).\footnote{This phenomenon is widely
labeled `common envelope evolution', but should not be confused with the
evolution of contact binaries within a \emph{quasistatic} common envelope.} 
Realistic three-dimensional models of common envelope evolution, including
radiative transfer, from onset to completion remain an unrealized dream. 
Rather, population synthesis models (for the distribution of properties of an
ensemble of coeval binaries) rely upon simple energy arguments plus an
efficiency parameter to estimate the outcomes of common envelope evolution
(e.g., de Kool 1992; Kolb 1993; Politano 1996). 
Even so, evolutionary analyses of the origins of known close double white
dwarfs frequently demand efficiencies greater than unity, an unmistakable sign
of deficiencies in even simple accounts of the energy budgets of these
binaries (see, e.g., Iben \& Livio 1993).

Certain types of binaries (for example, massive X-ray binaries,
symbiotic stars, and barium stars) are fueled or created through wind
accretion.  Here as well, three-dimensional radiation-hydrodynamical models
are needed for realistic treatments of mass and angular momentum accretion. 

Finally, while the broad effects of supernova kicks on the dynamics
of the binaries in which they occur have been explored (Sutantyo 1978; Hills
1983; Brandt \& Podsiadlowski 1995; Kalogera 1996), the dependence of the
magnitude and direction of those kicks on the history, mass, and orbit of the
exploding star remains unknown, but potentially of critical importance to the
survival or disruption of those binaries.

\subsubsection{New issues}

To this litany of unsolved problems in close binary evolution must now be
added several which are unique to a dense stellar environment, or at any rate
assume far greater significance there.

Close tidal encounters or mergers
can impart to the component stars or merger product rotational angular momenta
which would have been impossibly large for a main sequence star to
support.\footnote{This circumstance can arise in ordinary binary mass transfer
as well, although an argument can be made from the survival of Algol-type
binaries that tidal torques must in those cases be capable of mitigating the
concentration of angular momentum in the accreting star.}  Theoretical models
are not yet capable of answering how a thermally-distended merger product may
be capable of shedding its excess angular momentum and relaxing to a normal,
pressure-supported (if still rapidly rotating) state.

Mergers and binary
mass transfer can also give rise to \emph{non-canonical} stars, by which we
mean stars with chemical profiles which \emph{never} occur among single stars
-- mass-losing stars with anomalously-large cores for their masses, or
mass-gaining stars with anomalously-small cores.  Such stars appear
regularly in binary mass transfer calculations, but no systematic survey of
their evolution exists, nor may one even be practical.  These are stars for
which it will likely become necessary to integrate on-demand stellar
evolutionary calculations into evolving cluster dynamical models.\footnote{An
efficient implementation of this strategy will, however, require the
capability of interpolating detailed interior models from a library of single
star models for a given mass and age as predicate for constructing the desired
non-canonical star; and likewise the ability to identify circumstances in
which the interior of a non-canonical star has converged so closely with that
of a canonical star that it no longer need be followed in detail.} 

Another issue arising uniquely in dense stellar clusters is the role of
small perturbations from passing stars -- energy exchanges too small to be of
consequence for cluster evolution -- which may nevertheless exert a profound
influence on binary evolution.  For example, mass transfer rates in
cataclysmic binaries scale exponentially with the ratio of the difference
between stellar and Roche lobe radii to the stellar atmospheric pressure scale
height (Ritter 1988), a ratio which is typically of order $10^{-4}$. 
Fractional
perturbations to the orbital separation of this magnitude or higher could
profoundly effect the outburst behavior of those variables, or even
conceivably drive the white dwarf to evolve back toward the giant branch,
triggering a new common envelope phase of evolution.  Existing detailed models
of close binary mass transfer rest on the stability of mass transfer to
small-amplitude perturbations, but the stability of those models to
large-amplitude perturbations remains unexplored territory.

Among binaries
evolving in isolation, theory generally predicts that (post-supernova binaries
excepted) any orbital eccentricity will be tidally damped to insignificance
before Roche lobe overflow actually occurs (see, e.g., Zahn 1992; Tassoul \&
Tassoul 1996); but that circumstance certainly
cannot hold among the perturbed systems just described, in which the general
problem of mass transfer in eccentric binaries must be revisited.  Physically
sound solutions to all of these problems will require a more fundamental
understanding of the sources and properties of stellar viscosity than is now
at hand.

\subsubsection{Prospectus}

The added dimensionality of binary and merger evolutionary problems
effectively precludes the practicality of a library look-up approach as
advocated above for single star evolution.  The uncertainties in binary
evolution, particularly in the all-important mass and angular momentum loss
rates, are so great as to vitiate any attempt at present to build such a
library.  Rather, the most practical approach seems clearly a recipe-based
formalism, as is commonly used in population synthesis studies of binary
evolution, e.g. in the \textsc{bse} and \textsc{seba} codes mentioned in
Section 5.1
above.  This approach uses relatively simple, approximately formulae to
describe the outcome of mass transfer in a given situation.  These recipes
have been built up piecemeal from a qualitative understanding of the criteria
which dictate which evolutionary path a given binary will follow.  With no
systematic survey of close binary evolution in all three major dimensions
(mass, mass ratio, and orbital separation) yet practical, major gaps
inevitably remain in these prescriptions, but they have the virtues of extreme
speed; and for any single binary, the uncertainties in its outcome state are
in most circumstances dominated by the parametrization of mass and angular
momentum losses, which afflict detailed models and recipes alike.

Despite the bleak perspective offered above on the current state and
capabilities of binary or merger evolutionary models, one should not lose
sight of the fact that the duration of intense interaction in those models is
typically extremely brief, compared with cluster dynamical time scales.  Where
that is not the case (mass transfer driven on a nuclear or angular
momentum loss -- magnetic stellar wind or gravitational radiation -- time
scale), the donor star is in thermal equilibrium, and can (excepting
non-canonical stars) be well-approximated by an appropriate model from a
single-star library of evolutionary models.  Mass transfer rates and
evolutionary pathways can be derived implicitly from that library.  One
expects that, at a given instant in time, most stars in most binaries, whether
interacting or not, can be represented by members of a single star library.

\section{Stellar Dynamics and Stellar Evolution Interface}
\label{dynev}

\subsection{Single Stars, without Hydrodynamics}

During the workshop, we discussed the development of specifications
for the interfaces between different modules in simulations of dense
stellar systems.  As a concrete example, we focused first on the
simplest interface, that between stellar dynamics and stellar
evolution, without using hydrodynamics, and without allowing any
binaries.  Such an interface could be used in an $N$-body program
where single stars can collide and merge, while signaling the stellar
evolution counterparts of two merging stars to construct a new model
for the merger product.

In specifying such an interface, we do not want to make any assumption
about the computational processes that may take place at either side
of the interface.  The stellar evolution information may be provided
from look-up tables or fitting formulae based on sets of evolutionary
tracks that have been computed earlier, or it may be provided by an
actual stellar evolution code running in real time.  The stellar
dynamics information can come from an actual $N$-body code, or it can
come from a simple toy model that effectively produces a population
synthesis of distributions of colliding stars without any `live'
dynamics involved.  In all these and other cases, the interface should
not care what is happening at either side, as long as the specifications
for the interface are obeyed.

There are at least three quite different ways to specify an interface
between evolution and dynamics:

\begin{enumerate}

\item
{\it Minimal Interface.}
The stellar dynamics code can ask the stellar evolution code to take
one step forward.  This type of handshaking places the least demand
on the stellar evolution side of things, since everything is driven
from the stellar dynamics side.

\item
{\it Multi-Criterion Interface.}
The stellar dynamics code can ask the stellar evolution code to
proceed for a specified increase in time, as long as a number of
criteria are met (no unacceptably large changes in important physical
variables).  This gives the stellar dynamics more control over the
situation.  It requires some additional code to be written to steer
the stellar evolution code, but nothing very complicated.

\item
{\it Maximal Interface.}
The other extreme would be for the stellar evolution part of the code
to compute the complete future evolution of every new star created.
This may involve reading a precomputed table in the case of a star that starts
on the main sequence, or it may entail the production of a new table
by running a stellar evolution code, in the case of a merger product.

\end{enumerate}

While the first approach may be convenient, in that it requires the
least amount of changes to existing stellar evolution codes, such a
specification violates our requirement of a black-box approach.  It is
not clear what it would mean to ask a table look-up implementation of
stellar evolution to `take a next step'; nor can one ask a fitting
function to take a step.

In the second choice of interface there is a danger that the time
steps requested are unnaturally small from a stellar evolution point
of view, leading perhaps to unacceptable round-off error.  However,
such problems can be easily anticipated, for example by letting a star
take an evolutionary step only when the accumulated time increments
become comparable to its own natural integration time step.

In case 3), the stellar evolution module will provide a complete
future history of each new star, and will make it available to the
stellar dynamics side.  This will make it possible for the dynamics to
request the value of any physical quantity at any current time or any
predicted time in the future.  The main drawback is that too much
computer time may be used in the complicated late phases of unusual
stars, when those stars themselves may well merge again with another
star before reaching those stages.  This may not be a problem as long
as the majority of such stars do end their life without further
significant perturbations.

\subsection{Interface Function Specifications: an Example}

Below we describe in some detail what a multi-criterion interface could
look like.  Note that the specifications given there are only a first
step toward the simplest possible case.  A real specification should
include a treatment of binaries as separate computational objects,
which are far more complicated than those corresponding to single
stars.  And in any case, in practice we will wait to formalize the
interface until we have developed some experience with both the
single-star and the binary cases.  We may want to start with an alpha
specification, then a beta specification, and then freeze the
specification in a public version, in the sense that future interfaces
will have to at least respect the requirements listed in the public
version.

Here is a wish list from the stellar dynamics point of view.  We would
like to give an $N$-body program access to the following functions,
which can be considered as a type of library function call:

\begin{itemize}

\item[1)] a star creation function

\item[2)] a star evolution function

\item[3)] a star destruction function

\item[4)] a function providing the mass of a star

\item[5)] a function providing the radius of a star

\item[6)] a function providing the current time for a star

\item[7)] a function providing the total helium fraction of a star

\item[8)] a function providing the total metallicity of a star

\end{itemize}

For concreteness, we will write the specifications for these functions
in the form of desiderata for functions written in Fortran.  Similar
specifications can be prescribed for other languages, such as C or C++,
and in many cases the interface will be used to connect two modules
that may be written, say, in Fortran and C++.

\noindent
The function

\begin{verbatim}
    integer function
        CreateStar(M, Y, Z)
\end{verbatim}

\noindent
   accepts as arguments {\tt real*8} variables for the initial mass, the
   helium abundance and metallicity of a star created at the zero age
   main sequence.  The return value is a unique integer that acts as
   the identifier for the particular star that has been created.  A
   negative return value will signal an error (e.g. not enough storage
   left; unreasonable initial conditions provided; or some other internal
   error in the stellar evolution module).  The units for the variables are:

\begin{itemize}

\item[$M$] in solar masses

\item[$Y$] helium abundance fraction by weight;\\ $0 \le Y \le 1$

\item[$Z$] metallicity abundance by weight;\\ $0 \le Y + Z \le 1$

\end{itemize}

\noindent
   The function

\begin{verbatim}
    real*8 function
        EvolveStar(id, dtmax, dMmax,
                   dRmax, dYmax, dZmax)
\end{verbatim}

\noindent
   accepts as first argument an integer variable for the identifier
   {\tt id}, followed by five {\tt real*8} variables that determine
   halting criteria.  The stellar evolution code will start evolving
   the star, from the current time $t_{now}$, at which the mass,
   radius, and compositions are $M_{now}$, $R_{now}$, $Y_{now}$,
   $Z_{now}$.  The code will stop as soon as one of the following
   halting criteria is satisfied:

\begin{itemize}

\item[if]  the time $t \ge t_{now} + dt $

\item[if]  the mass $M$ obeys $| M - M_{now} | > dM_{max}$

\item[if]  the radius $R$ obeys $| R - R_{now} | > dR_{max}$

\item[if]  the helium fraction $Y$ obeys $| Y - Y_{now} | > dY_{max}$

\item[if]  the metallicity $Z$ obeys $| Z - Z_{now} | > dZ_{max}$

\end{itemize}

   The function returns the new time $t$; a negative value for $t$ indicates
   an error condition.  The additional units used here are:

\begin{itemize}

\item[$t$] in millions of years

\item[$R$] in solar radii

\end{itemize}

\noindent
   The function

\begin{verbatim}
    integer function
        DestroyStar(id)
\end{verbatim}

\noindent
   accepts an integer {\tt id}, the identifier for the star that should
   be destroyed.  The function will remove that star, freeing up the
   memory assigned to it.  Successful completion will be indicated by
   returning a positive or zero integer; a negative integer will
   indicate an error condition.

\noindent
   The function

\begin{verbatim}
      real*8 getMass(id)
\end{verbatim}

\noindent
accepts an integer {\tt id}, and returns the value for the mass of the
corresponding star.

\noindent
   Similarly the functions

\begin{verbatim}
    real*8 getRadius(id)
    real*8 getTime(id)
    real*8 getY(id)
    real*8 getZ(id)
\end{verbatim}

\noindent
accept an integer {\tt id}, and return the values for the stellar
radius, current, helium fraction, and metallicity, respectively.

\subsection{Implementation in an $N$-Body Program}

The above eight functions suffice for the inclusion of stellar
evolution into a stellar dynamics program, at least for the case
where no interacting binaries are present.  Everything else will
be controlled and determined directly by the SD program.  For
example, if the SD program determines that the distance between
two stars has become smaller than the sum of the radii of the two
stars, the SD program can merge the two stars (with identifiers
{\tt id1} and {\tt id2}) as follows:

\begin{verbatim}
    real*8 m1,m2,newmass,newY,newZ
    integer newstar
        ...
    m1 = getMass(id1)
    m2 = getMass(id2)
    newmass = m1+m2
    newY = (getY(id1)*m1
            + getY(id2)*m2)/newmass
    newZ = (getZ(id1)*m1
            + getZ(id2)*m2)/newmass
    newstar = 
        CreateStar(newmass,newY,newZ)
\end{verbatim}

\noindent
    This code fragment creates a homogeneous ZAMS star from the
    matter obtained by adding the previous two stars.  This
    procedures assumes complete mixing and ignores transient effects
    that will die out during a thermal time scale, such as an increase
    in radius due to shock heating.  Other complications, such as a
    possibly rapid rotation after the merger are neglected as well.

With a black box approach, it is vital to test for possible errors,
since you have no idea what is going on internally.  The right
defensive programming approach would be to let the lines above be
followed by error checks, with appropriate actions (here indicated by
$...$ for each particular type of error):

\begin{verbatim}
    if (newstar .lt. 0) then
        ...
    endif
    if (DestroyStar(id1) .lt. 0) then
        ...
    endif
    if (DestroyStar(id2) .lt. 0) then
        ...
    endif
\end{verbatim}

\noindent
    For any serious production runs, the above treatment will need to
    be extended to model binaries as well.  Such an extension may well
    imply modifications of the above simple treatment.  Therefore, we
    do not intend our presentation here to be definitive in any way.
    Indeed, even for single stars, we may want to extend the above
    interface, for example by including the possibility of a star
    receiving a kick-velocity at the time of a supernova explosion.
    The only requirement will be that an agreed-upon future version of
    the above specification, when adapted as a standard, should remain
    valid in later versions that will be upwards compatible with that
    standard.

\section{Stellar Hydrodynamics}
\label{hydrodynamics}

\subsection{Smoothed Particle Hydrodynamics}
\label{SPH}

Hydrodynamic interactions such as collisions and mergers can
strongly affect the overall energy budget of a cluster and even
alter the timing of important dynamical phases such as core
collapse. Furthermore, stellar collisions and close encounters are
believed to produce a number of non-canonical objects, including
blue stragglers, low-mass X-ray binaries, recycled pulsars, double
neutron star systems, cataclysmic variables and contact binaries.
As discussed in \S\ref{evolution}, these stars and systems are
among the most challenging to model, and they are also among the
most interesting observational markers.  Predicting their numbers,
distributions and other observable characteristics is essential
for detailed comparisons with observations.

In galactic nuclei, collisions are very energetic events that
typically result in two unbound stars that have suffered mass loss
(a kind of ``fly-by").  Mergers are actually a rare outcome, as
collisions with small impact parameters often result in complete
destruction of the parent stars \cite{fre02a}. In globular
clusters, the velocity dispersion is less than the escape velocity
from the surface of a parent main sequence star, and therefore
mergers are much more likely.  In any case, the structure and
chemical composition profiles of a collision product are clearly
of central importance, because they determine its observable
properties and evolutionary track in a color magnitude diagram
(e.g.\ Sills \& Bailyn 1999).

Three-dimensional hydrodynamic simulation is one means to study
stellar collisions and to determine the trajectories and interior
profiles of the resulting products.  Mostly using the Smoothed
Particle Hydrodynamics (SPH) method, numerous scenarios of stellar
collisions and interactions have been simulated in recent years,
including, for example, collisions between two main sequence stars
\cite{ben87,lai93,lom96,oue98,san97,sil97,sil01,fre02a},
collisions between a giant star and compact object \cite{ras91},
collisions during three- and four-body interactions
\cite{goo91,dav94}, and common envelope systems
\cite{ras96,ter94,ter95,san98,san00}.

SPH is a Lagrangian technique in which the system is broken up
into a large number of fluid particles whose positions and
velocities are integrated forward in time according to
hydrodynamic and self-gravitational forces. Local densities and
hydrodynamic forces at each particle position are calculated by a
kernel estimation that involves summing over nearest neighbors.
For an overview of the basic SPH equations, see, for example,
Monaghan(1992) or Rasio and Lombardi (1999).

The so-called entropic variable $A\equiv P/\rho^{\Gamma}$ turns
out to be critical for understanding the physics of mergers and
therefore the results of SPH simulations. Here $P$ is pressure,
$\rho$ is density, and $\Gamma$ is the adiabatic index of the gas
(assumed here to be constant). Given the importance of $A$, we
will first discuss this quantity in the context of single,
isolated stars.  It is straightforward to show analytically that
the condition ${\rm d}A/{\rm d}r>0$ is equivalent to the usual
Ledoux criterion for convective stability of a non-rotating star
\cite{lom96}.  The basic idea can be seen by considering a small
fluid element inside a star in dynamical equilibrium.  If this
element is perturbed outward adiabatically (that is, with constant
$A$), then it will sink back toward equilibrium only if its new
density is larger than that of its new environment. Because
pressure equilibrium between the element and its immediate
environment is established nearly instantaneously, the ratio of
densities satisfies
\begin{displaymath}
\rho_{\rm ele}/\rho_{\rm env}=
(A_{\rm ele}/A_{\rm env})^{-1/\Gamma}.
\end{displaymath}
Therefore, a
fluid element with a lower $A$ than its new surroundings will sink
back down toward the equilibrium position. Likewise, if an
inwardly perturbed fluid element has a larger $A$ than its new
environment, buoyancy will push the element outwards, back toward
equilibrium. As a result, a stable stratification of fluid
requires that the entropic variable $A$ increase outward: ${\rm
d}A/{\rm d}r>0$. In such a star, a perturbed element will
experience restoring forces that cause it to oscillate about its
equilibrium position.  For a detailed discussion of the stability
conditions within rotating stars, see \S 7.3 of \citeasnoun{tas78}
or \citeasnoun{tas00}.  In practice, SPH calculations show that,
even in rapidly rotating stars, fluid distributes itself in such a
way that the entropic variable $A$ increases outwards.

For a merger of stars, SPH simulations reveal that fluid
elements with low values of $A$ do indeed sink to the bottom of a
gravitational potential well, and the $A$ profile of a merger
product in stable dynamical equilibrium increases radially
outwards. Because this $A$ profile is typically steep, especially
in the outermost layers, collision products, in contrast to normal
pre-main sequence stars, do not develop convective envelopes. An
additional consequence of having the fluid stratify itself
according to $A$ is that parent stars are not thoroughly mixed
during collisions; instead, strong chemical composition gradients
are present even in the final configuration. The stellar evolution
of collision products therefore can depart significantly from that
of normal stars that begin their lives as chemically homogeneous,
``zero-age'' main sequence stars.

Because the quantity $A$ depends directly upon the chemical
composition and the entropy, it is conserved for each fluid
particle during gentle, adiabatic processes. During a collision,
the entropic variable $A$ of a fluid element can increase due to
shock heating. However, at least in open, globular and young
compact star clusters, the relative impact speed of two stars is
comparable to the speed of sound in these parents: both speeds are
of order $(GM/R)^{1/2}$, where $G$ is Newton's gravitational
constant, and $M$ and $R$ are respectively the mass and length
scales of a parent star. Consequently, the resulting shocks have
Mach numbers of order unity and shock heating is relatively weak.
Therefore, to a reasonable approximation, a fluid element
maintains a constant $A$ throughout a collision.  Hyperbolic
collisions, appropriate in galactic nuclei, do result in
significantly more shock heating; however, the brunt of the shocks
are absorbed by what becomes the ejected mass, thereby shielding
the cores of the parent stars at least in less extreme cases.

\subsection{Generating Collision Product Models, Quickly}
\label{fasthydro}

A substantial fraction of the stars in a cluster will experience a
collision sometime during their lifetimes.  The direct integration
of low resolution SPH calculations into a cluster evolution code
may allow the modeling of such events in the not very far future.
However, a single high resolution hydrodynamic simulation can
typically take hundreds or even thousands of hours to complete.
Therefore, the excessive computing time required of hydrodynamics
simulations makes it extremely impractical, if not currently
impossible, to couple them directly to stellar dynamics
calculations.

One solution, taken by \citeasnoun{fre02a}, is to calculate first
an extensive set of hydrodynamics simulations, varying the parent
stars, as well as the eccentricity and periastron separation of
their initial orbit.  The SPH database of Freitag \& Benz treats
all types of hyperbolic collisions between main sequence stars:
mergers, fly-bys and cases of complete destruction.  The tremendous
amount of parameter space surveyed precludes having high enough
resolution to determine in detail the structure and composition
profiles of the collision products for all cases; however,
critical quantities such as mass loss and final orbital elements
are indeed determined accurately.  By interpolating among these
hydrodynamics results, \citeasnoun{fre02b} have successfully
integrated collisions into a Monte Carlo star cluster code,
yielding the most realistic treatment ever of stellar collisions
in a stellar dynamics code.

It should be noted that, even without relying directly on SPH
results, certain aspects of collisions could be modeled in a
cluster simulation code using existing techniques. For example,
Fig. 3 of \citeasnoun{fre02a} shows that the mass loss in high
velocity collisions with relatively large impact parameters
[$d_{min}/(R_1+R_2) \gtrsim 0.5$] is surprisingly well predicted
by a simple method devised by Spitzer \& Saslaw (1966) and based
on conservation of momentum and energy. Such collisions are likely
to occur in a galactic nucleus near a massive black hole, as high
velocities quench focusing and make collisions with small impact
parameters rare.

As an approach for generating merger product models without
running hydrodynamics calculations, \citeasnoun{lom02} have
developed a method that calculates the structure and composition
profiles from simple algorithms based on conservation laws and a
basic qualitative understanding of the hydrodynamics. The
thermodynamic and chemical composition profiles of the simple
models, as well as their subsequent stellar evolution, agree very
well with those from the SPH models. Because the method takes only
a few seconds to generate a model on a typical workstation, it
becomes feasible to incorporate the effects of mergers in
dynamics simulations of globular clusters. The algorithms have
been implemented in an easy to use software package dubbed ``Make
Me A Star.''\footnote{\tt http://vassun.vassar.edu/$\sim$lombardi/mmas}

The underlying principle behind this method exploits the two
special properties of $A$ discussed in \S\ref{SPH}: Namely, the
entropic variable $A$ will (1) increase outward in a stable star
and (2) be approximately conserved during a collision.  Therefore,
to a good approximation, {\it the distribution of fluid in a
merger product can be determined simply by sorting the fluid
from both parent stars in order of increasing $A$}: the lowest $A$
fluid from the parent stars is placed at the core of the merger
product and is surrounded by shells with increasingly higher $A$.
This treatment is further improved upon by modeling the shock
heating, hydrodynamic mixing, mass ejection, and angular momentum
distribution with physically motivated fitting formulas calibrated
from the results of SPH simulations. Although the algorithms
currently are capable of treating only parabolic collisions
between stars obeying an ideal gas equation of state, the method
and code have been developed with the intent of ultimately
generalizing to other collision scenarios.

\section{Stellar Dynamics, Hydrodynamics and Stellar Evolution Interfaces}
\label{dynhyev}

\subsection{Including Hydrodynamics}

Here we discuss how the communication between a stellar dynamics
(SD) and stellar evolution (SE) module, presented in
\S\ref{dynev}, can be extended to include an additional interface
with a stellar hydrodynamics (SH) module.
%
%
The SD module will continue to be the scheduler and manager,
passing only  the minimum amount of data necessary to any SE or SH
routine. However, for hydrodynamic processes to be modeled, we
must now also allow for the storage and retrieval of stellar
structure and composition profiles.  The general purpose of the SH
module is to take such profiles for parent stars, and return
profiles for newly created collision product(s).  Rotation of
collision products is neglected in this simple interface, but could
be treated by also including a profile for the specific angular
momentum.

SH is likely to play an important role in single-binary and
binary-binary  interactions.  In such cases, a merger product
created in an initial collision  will have a greatly enhanced
collisional cross-section due to the shock  heating of fluid.  A
second (and even a third) collision can likely result  before the
first merger product significantly contracts as it thermally
relaxes. Communication among the various modules is therefore
high: the SD module tracks the stellar trajectories, the SH module
generates collision product models, and  the SE module evolves the
structure of these models.  The simple SH module that we discuss
below would treat each successive collision separately,
neglecting tidal forces from nearby stars and the slight
possibility that a third star could collide or strongly interact
while the first two parent stars are in the process of colliding.

\subsection{Routines Provided by the Stellar Hydrodynamics Module}
\label{shmodule}

There are two main routines supplied by the SH module: (1) a
stellar collision function that determines what happens during a
collision, and (2) a subroutine that returns the structure and
composition profiles, as well as the position and velocity, of any
collision products.  For concreteness, we
will write the specifications for these functions in Fortran (taking
some liberties with indentation and continuation lines).

The function

\begin{verbatim}
    integer function
        CollideStars(r,v,
            mProfile1,rProfile1,
            PProfile1,rhoProfile1,
            chemicalProfiles1,
            mProfile2,rProfile2,
            PProfile2,rhoProfile2,
            chemicalProfiles2,
            numShells1,numShells2,
            numChemicals)
\end{verbatim}

accepts input arguments declared as follows.

\begin{verbatim}
    real*8 r(3), v(3)
    integer numShells1,numShells2,
            numChemicals
    real*8 mProfile1(numShells1),
           rProfile1(numShells1),
           PProfile1(numShells1),
           rhoProfile1(numShells1),
           mProfile1(numShells2),
           rProfile2(numShells2),
           PProfile2(numShells2),
           rhoProfile2(numShells2),
           chemicalProfiles1(numShells1,
                             numChemicals),
           chemicalProfiles2(numShells2,
                             numChemicals)
\end{verbatim}

The arrays {\tt r} and {\tt v} specify, respectively, the relative
position and velocity of parent star 2 with respect to parent star 1,
in Cartesian coordinates.  The integers {\tt numShells1} and {\tt
numShells2} give the number of shells in which the structure and
chemical composition profiles are stored for stars 1 and 2,
respectively.  The integer {\tt numChemicals} gives the number of
different chemical species that are being considered, presumed to be
the same for each parent star. 

The arrays {\tt mProfile1}, {\tt rProfile1}, {\tt PProfile1} and {\tt
rhoProfile1} specify the structure of parent star 1 by specifying its
enclosed mass, radial, pressure and density profiles, respectively.
That is, {\tt mProfile1(i)}, {\tt PProfile1(i)} and {\tt
rhoProfile1(i)} give the enclosed mass, pressure and density,
respectively, in star 1 at a spherical shell of radius {\tt
rProfile(i)}, for any integer {\tt i} in the range from 1 to {\tt
numShells1}. These arrays contain redundant information that could
also have been obtained by integrating $dm=4\pi r^2 \rho dr$ or the
equation of hydrostatic equilibrium (if appropriate); nevertheless, as
a matter of convenience it is useful to have all four arrays
available.

The array element {\tt chemicalProfiles(i,j)} gives the fractional
composition, by weight, of chemical species number {\tt j} in shell
{\tt i}. Similarly, the arrays {\tt mProfile2}, {\tt rProfile2}, {\tt
PProfile2}, {\tt rhoProfile2} and {\tt chemicalProfiles2} specify the
structure and composition profiles of parent star 2.  The last element
of the radial profile arrays, {\tt rProfile1(numShells1)} and {\tt
rProfile2(numShells2)}, are taken as the stellar radii of the parents.
Likewise, {\tt mProfile1(numShells1)} and {\tt mProfile2(numShells2)}
are the masses of the parent stars.  All values are stored in cgs
units.  Chemical composition profiles are dimensionless, as they
represent the fractional abundance by mass.

The function {\tt CollideStars} returns the number of collision products
generated. For parabolic and weakly hyperbolic encounters the
returned value will often be  1, but it could also be 2 (with the
two stars having new mass and internal  profiles).  For strongly
hyperbolic encounters, the returned integer is 0  when the stars
are destroyed by the collision, and possibly, at least in
principle, larger than 2 if multiple stars collapse out of a
remnant gas cloud after a catastrophic collision.  A returned
value that is negative will signal an error condition (e.g., the
input relative separation, velocity  and stellar radii not being
consistent with a close interaction).

Internally this routine will generate structure and chemical
composition  profiles for the collision product(s).  This could be
done in any number of ways, for example by actually running an SPH
simulation, by interpolating SPH results, or with simple recipes
or fitting formula (see \S\ref{hydrodynamics}).  These profiles
will be stored in memory until the next call to {\tt CollideStars}, and can
be retrieved in  the meantime through the subroutine {\tt
getProduct}.

When called after the function {\tt CollideStars}, the subroutine

\begin{verbatim}
    getProduct(rProduct,vProduct,
        mProfile,rProfile,PProfile,
        rhoProfile,chemicalProfiles,
        numShells)
\end{verbatim}

\noindent returns the position and velocity, as well as the
structure and composition profiles, of the collision product(s)
generated in the most recent call to {\tt CollideStars}.  The
returned arrays {\tt rProduct} and {\tt vProduct} specify,
respectively, the relative position and velocity of a collision
product with respect to the input (pre-collision) position and
velocity of parent star 1, in Cartesian coordinates.  The enclosed
mass, radius, pressure, density and chemical composition profiles
are returned as the {\tt real*8} arrays {\tt mProfile}, {\tt
rProfile}, {\tt PProfile}, {\tt rhoProfile} and {\tt
chemicalProfiles}, respectively.

The first call to {\tt getProduct} yields the profiles for the first
collision product; the second call is for the second collision
product, etc.  Also returned, as the final argument to the subroutine,
is the integer {\tt numShells} specifying the number of shells in the
structure and composition arrays.  The {\tt chemicalProfiles} array is
two dimensional with the second argument running from 1 up to {\tt
numChemicals}, automatically set to the same number of chemical
species being considered as in the parent stars.

The declaration of the arguments for {\tt getProduct} is
therefore as follows.

\begin{verbatim}
    integer numShells
    real*8 rProduct(3), vProduct(3)
    real*8 mProfile(numShells),
           rProfile(numShells),
           PProfile(numShells),
           rhoProfile(numShells),
           chemicalProfiles(numShells,
                            numChemicals)
\end{verbatim}

\subsection{Routines Provided by the Stellar Evolution Module}


Even with the inclusion of a SH module,  the {\tt EvolveStar},
{\tt DestroyStar}, {\tt getMass}, {\tt getRadius} and {\tt
getTime} functions would not need to be modified from their
versions described in \S\ref{dynev}.  However, the {\tt
CreateStar} function does need to be generalized to allow for the
creation of stars with arbitrary structure and composition
profiles.

The function
\begin{verbatim}
    integer CreateStarFromProfiles(
        mProfile,rProfile,PProfile,
        rhoProfile,chemicalProfiles,
        nucleonNum,
        numShells,numChemicals)
\end{verbatim}

\noindent fills this role, accepting as arguments {\tt real*8}
arrays for the enclosed mass, radius,  pressure, density and
chemical composition profiles, in that order.  The intent is that
these arrays will have been generated through calls to {\tt
CollideStars} and {\tt getProduct}. The two dimensional {\tt
real*8} array {\tt nucleonNum} specifies the chemical species
being considered, with {\tt nucleonNum1(1,j)} and {\tt
nucleonNum1(2,j)} giving the number of protons and neutrons,
respectively, for species {\tt j}.

The final two input arguments are the integers {\tt numShells} and
{\tt numChemicals} that, respectively, specify the number of shells
and chemical species represented by these arrays.  As in the {\tt
CreateStar} function of \S\ref{dynev}, the return value is a unique
integer that acts as the identifier for the particular star model that
has been created, with a negative return value signaling an error.

With the introduction of an SH module, another new routine
required from the SE module is a subroutine providing the current
profiles of a star:

\begin{verbatim}
    getProfiles(id,mProfile,rProfile,
        PProfile,rhoProfile,
        chemicalProfiles,numShells,
        numChemicals)
\end{verbatim}

\noindent
This subroutine accepts an integer {\tt id} identifying a particular
star model.  The output arrays contain the same type of information as
the arrays returned by {\tt getProduct} (see \S\ref{shmodule}).
The chemical profiles returned in {\tt chemicalProfiles} are for the
same species, in the same order, as when the star was created.

\subsection{Implementation in the Stellar Dynamics Code}

The following code fragment will collide two parent star models
with id  numbers {\tt id1}  and {\tt id2}.  When no errors result,
the two parent stars will be  destroyed, and any collision product
models will be created.  Error conditions would be handled in
the portions of the code represented by ``...''

\begin{verbatim}
    getProfiles(id1,mProfile1,rProfile1,
                PProfile1,rhoProfile1,
                chemicalProfiles1,
                numShells1,numChemicals1)
    getProfiles(id2,mProfile2,rProfile2,
                PProfile2,rhoProfile2,
                chemicalProfiles2,
                numShells2,numChemicals2)
    numproducts=
        CollideStars(r,v, mProfile1,
                     rProfile1,PProfile1,
                     rhoProfile1,
                     chemicalProfiles1,
                     mProfile2,rProfile2,
                     PProfile2,rhoProfile2,
                     chemicalProfiles2,
                     numShells1,numShells2,
                     numChemicals)
    if(numproducts.ge.0) then
        if (DestroyStar(id1).lt.0) ...
        if (DestroyStar(id2).lt.0) ...
        do i=1,numproducts
            call getProduct(rProduct(1,i),
                vProduct(1,i),
                mProfile,rProfile,
                PProfile,rhoProfile,
                chemicalProfiles,numShells)
            idarray(i)=CreateStar(
                mProfile,rProfile,PProfile,
                rhoProfile,chemicalProfiles,
                nucleonNum,
                numShells,numChemicals)
            if(idarray(i).lt.0) ...
        enddo
    else
         ...
    endif
\end{verbatim}

Even if one wanted to make the perhaps crude approximation that
a merger product somehow became chemically homogeneous on a
short timescale, the composition information provided by {\tt
getProduct} can still be nontrivial: the ejected
mass in collisions comes preferentially from the outermost  layers
of the stars, and therefore the total composition fractions in a
collision product are not exactly a simple mass average of the total
fractions in the parent stars.

%
%
%

\section{Formation of Stars and Stellar Systems}
\label{sec:initial-conditions}

To be able to follow the entire life cycle of stellar systems, we need
to understand how stars form, and in particular, how stars form in
dense aggregates and clusters. This knowledge allows us to define
astrophysically relevant initial conditions for the in-detail
investigation of the subsequent dynamical evolution of the cluster --
the main aim of the MODEST collaboration. We briefly review here
scope and limitations of the current numerical models of
molecular cloud fragmentation and star cluster formation and then add
some further more general considerations.

\subsection{Turbulent Fragmentation and the Formation of Stellar
  Clusters}
\label{subsec:turb-frag}
Careful stellar population analysis indicates that most stars in the
Milky Way (of order of 90\%) form in open clusters with a few hundred
member stars. Rich stellar clusters with several thousands to ten
thousands of stars account for most of the remaining stars
\cite{AM01}. Very rich stellar clusters with several hundred thousand
or millions of stars (e.g.\ globular clusters) are extremely rare, and
contribute only a very small fraction of the entire stellar population
of a galaxy.

To our current understanding, all stars are born in turbulent interstellar
clouds of molecular hydrogen.  In the so called ``standard'' theory of star
formation, stars build up from the inside-out collapse of singular isothermal
spheres, which are generally assumed to result from the quasistatic
contraction of magnetically supported cloud cores due to ambipolar diffusion
(Shu 1977, Shu {\it et al.} 1987).  This picture is able to reproduce certain
observed features of protostars, however, clearly fails to explain {\em all}
known properties of protostars and star forming regions. Furthermore, some of
its key assumptions may not be met in typical molecular clouds. See, e.g.,
Whitworth {\it et al.} (1996), Nakano (1998), or Andr{\'e} {\it et al.} (2000)
for a critical discussion, and Crutcher (1999) and Bourke {\it et al.} 2001
for a compilation of magnetic fields determinations in molecular cloud cores.
This theory needs to be expanded or replaced by a more dynamical point of view,
realistically taking into account interstellar turbulence, which typically is
supersonic and super-Alfv{\'e}nic.  Supersonic turbulence establishes a
complex network of interacting shocks in molecular cloud, where converging
shock fronts generate clumps of high density. The density enhancement may be
large enough for the fluctuations to become gravitationally unstable and
collapse (e.g., Elmegreen 1993, Padoan 1995, Padoan \& Nordlund 1999). This
happens when the local Jeans length becomes smaller than the size of the
fluctuation.  However, fluctuations in turbulent velocity fields are highly
transient.  The random flow that creates local density enhancements can
disperse them again.  For local collapse to actually result in the formation
of stars, individual gravitationally unstable shock-generated density
fluctuations must collapse to sufficiently high densities on time scales
shorter than the typical time interval between two successive shock passages.
Only then are they able to `decouple' from the ambient flow and survive
subsequent shock interactions.  The shorter the time between shock passages,
the less likely these fluctuations are to survive.  Hence, the efficiency of
protostellar core formation, the growth rates and final masses of the
protostars, essentially all properties of the nascent star cluster strongly
depend on the intricate interplay between gravity on the one hand side and the
turbulent velocity field in the cloud on the other.

Clusters of stars build up in molecular cloud regions where
self-gravity overwhelms turbulence (see, e.g., Clarke {\it et al.}
2000, Elmegreen {\it et al.} 2000), either because the region is
compressed by a large-scale shock (e.g., Klessen {\it et al.} 2000, or
Heitsch {\it et al.} 2001), or because interstellar turbulence is not
replenished and decays on short timescales (Mac~Low {\it et al.}\ 
1998, Stone {\it et al.} 1998, Mac Low 1999).  Once individual gas
clumps become gravitationally unstable within the star forming region,
they begin to collapse.  The gas density increases and a hydrostatic
protostellar object forms in the center of the collapsing core.  In
dense clusters, collapsing gas clumps may merge, producing new clumps
that then contain multiple protostars.  Dynamical interactions are
common, close encounters occur frequently and will drastically alter
the trajectories, thus changing the accretion rates.  This has
important consequences for the final stellar mass distribution
(Bonnell {\it et al.} 2001a,b, Klessen 2001a,b).  Already in their
infancy, i.e.\ already in the deeply embedded phase, stellar clusters
are strongly influenced by collisional dynamics.  Turbulent molecular
cloud fragmentation, competitive accretion, and protostellar
interaction, all are highly stochastic processes.  In essence, a
comprehensive theory of star formation thus needs to be a statistical
theory.

In sufficiently populous clusters O stars may form.  Their intense UV
radiation photoionizes the surrounding molecular cloud region, and together
with their strong winds lead to the rapid expulsion of the residual gas on
timescales typically faster than the dynamical time from the cluster (e.g.,
Churchwell 1999). This leads to rapid dispersal of a large fraction of the
embedded cluster population (e.g., Kroupa {\it et al.} 2001, Boily \& Kroupa
2002).  The velocity dispersion of the expanding population is a function of
the binding energy of the embedded cluster, and in massive clusters it may
reach a few tens km/s.  Such kinematically hot components may lead to the
thickening of thin galactic disks in those instances when a disk galaxy goes
through a star-burst phase.  The thick disk of the Milky Way galaxy, as well
as the hitherto not understood steep rise of the age--velocity dispersion
relation of solar-neighborhood stars, may be a direct consequence of such
processes. Star-cluster birth may therefore be a necessary ingredient if we
are to understand the structural and kinematical properties of galaxies
(Kroupa 2002).

The following is a list of results of star cluster formation
calculation that are of relevance in the context of MODEST, i.e.\ for
combining star cluster formation with star cluster evolution:

\begin{enumerate}
\item Star clusters form fast, on timescales of order of the crossing time
  (e.g., Ballesteros-Paredes {\it et al.} 1999, Klessen \& Burkert 2000, 2001,
  Elmegreen 2001).
\item Star clusters form with a considerable degree of substructure.
\item Star clusters form with  a very high initial
  binary fraction (larger than 60\%). This is consistent with inverse
  population synthesis models (Kroupa 1995).
\item The stellar mass spectrum predicted by turbulent cloud fragmentation in
  cluster forming regions is consistent with observational determinations of
  the IMF (Klessen 2001b, Padoan \& Nordlund 2002; IMF see Kroupa 2002) and
  extends down into the brown dwarf regime (Bate {\it et al.} 2002).
\item Massive stars begin to form first and are able to maintain a
  high accretion rate. This is because massive stars form from the
  most massive and densest gas clumps in the star forming region. As
  these clumps are dense, collapse progresses fast. And because they
  are very massive, they constitute a local minimum of the cluster
  potential and are able to attract the inflow of further gas. Massive
  protostars therefore experience high accretion rates over an
  extended period of time. Low mass stars form from low mass gas
  clumps and only benefit from a short period of peak accretion (e.g., Klessen
  2001a).
\item Star clusters are expected to form mass segregated. Massive
  stars form close to the cluster center, low mass stars are likely to
  form at large cluster radii. This is because massive clumps
  constitute the central region of the nascent cluster and low mass
  clumps are predominantly found at the outskirts. Altogether, one
  expects a radial dependence of the cluster IMF. The current star
  forming models, however, give only first hints of this effect (e.g., Klessen
  {\it et al.} 2000, Bate {\it et al.} 2002). Small number
  statistics do not allow for detailed predictions yet.
\item During its first few million years, a star cluster will contain
  a mixture of stars on the main sequence and stars still in the
  pre-main sequence phase. Very massive stars enter the main sequence
  already during the main accretion phase, while the pre main sequence
  contraction phase of low mass protostars may last for several
  $10^7\,$years (e.g.\ Palla \& Stahler 1999). 
\item Star formation likely is a self-regulated process. Bipolar
  outflows from young stars stir the gas in star forming regions,
  thus modulating the accretion efficiency. Radiation from young stars
  will heat the gas. If O or B stars form, they will ionize the gas in
  their surrounding and prevent further mass growth and star
  formation. The same holds for supernovae explosions, which will also
  blow away the cluster gas. Regardless of the mechanism the removal
  of the remaining cluster gas terminates the star formation process
  and determines the efficiency of star formation.
\end{enumerate} 

\subsection{Numerical Models of Clustered Star Formation }
\label{subsec:num-mod}
Most numerical calculations to describe molecular cloud fragmentation
and star cluster formation use SPH to solve the equations of
hydrodynamics that govern the dynamical evolution of gaseous clouds.
Owing to the stochastic nature of supersonic turbulence, it is not
known in advance where and when local collapse occurs.  SPH is the
method of choice because it is fully Lagrangian. The fluid is
represented by an ensemble of particles and flow quantities are
obtained by averaging over an appropriate subset of the SPH particles
(Benz 1990, Monaghan 1992).  The method is able to resolve large
density contrasts as particle are free to move and so naturally the
particle concentration increases in high-density regions. In addition,
one can introduce `sink' particles into SPH (e.g.\ Bate {\it et al.}
1995), which have the ability to accrete gas from their surroundings,
while keeping track of mass and linear and angular momentum. By
adequately replacing high-density protostars in the centers of
collapsing gas clumps with sink particles, one is able to follow the
dynamical evolution of the system over many free-fall times. This is
an essential ingredient for following the formation of stellar
clusters.

The first attempts to numerically model the formation of star clusters date
back to the late 1970's (e.g.\ Larson 1978). With the rapid increase of
computer power in recent years, more realistic calculations became possible.
Whitworth et al. (1995) and Bhattal et al. (1998) investigated in detail the
fragmentation of shocked interfaces of colliding molecular clumps into small
stellar systems.  Klessen {\it et al.} (1998), and Klessen \& Burkert (2000,
2001) studied the formation of stellar clusters from random Gaussian density
fluctuation in molecular clouds. Models that include molecular cloud
turbulence and consistently follow clustered star formation from turbulent
fragmentation have been presented by Klessen {\it et al.}  (2000), Klessen
(2001a,b) and Bate {\it et al.} (2002).  Focusing on the role of competitive
accretion and neglecting the processes that lead to the formation of
protostellar cores, Bonnell et al.\ (2001a,b) study the mass growth of
randomly placed accretion particles in simplified model clouds.  It should be
mentioned, that the majority of numerical studies of interstellar turbulence
and molecular cloud fragmentation are based on grid-based methods (for further
references, see V{\'a}zquez-Semadeni {\it et al.} 2000). These models are
conceptually more difficult to combine with star cluster evolution
calculations than the SPH models discussed here.

Star formation is an enormously complex process. It spans 20 orders of
magnitude in density (from molecular cloud cores to the stellar
interior) and 7 -- 8 decades in spatial scale (compare the pc-size
scales of molecular clouds with typical stellar radii of
$\sim 10^{11}\,$cm). And it involves a large number of physical processes.
An adequate treatment of star formation must not only take into
account gas dynamics and self-gravity, as most models do, but also
include heating and cooling processes, radiation transfer, magnetic
fields, chemical phase transitions and reactions, and feedback
processes from star formation itself. Star formation very likely is a
feedback regulated process.  Bi-polar outflows from young protostars
deposit energy and momentum in star forming regions, stellar winds
heat the remaining gas, UV radiation from massive stars may completely
ionize the cluster gas thus preventing further gas accretion and
terminating star formation. Supernovae explosions (e.g.\ from very
massive young stars, or from a nearby OB association) finally may
disrupt molecular clouds altogether, preventing further star formation
on scales of cloud as a whole. One needs to keep in mind, that most of
these processes are {\em not} included in the models of star cluster
formation. Some first attempts to model the effects of gas expulsion
on the subsequent dynamical evolution of young stellar clusters  are
reported by Geyer \& Burkert (2002) and also by Kroupa {\it et al.} 
(2001).

\subsection{Initial Conditions from Star Formation Calculations}
\label{subsec:init-cond-SF-calc}
In principle, it appears straightforward to adopt the results of
molecular cloud fragmentation and star forming calculations as
starting conditions of star cluster evolution models. First, one needs
to identify all protostars in the star formation model, then second,
determine their masses $m$, positions $\vec{r}$, and velocities
$\vec{v}$, maybe also their angular momentum (spin $\vec{j}$), and
finally, supply this list to the stellar dynamics code.

However, in practice the situation is not that simple. Before we
realistically apply results from star formation models to star cluster
evolution we have to address several inconsistencies of the methods.
The following gives a list of assumptions that need to be introduced
to be able to combine both methods.

\begin{enumerate}
\item {\em Gas removal:}
  Because star forming calculations typically do not treat
  protostellar feedback and gas removal, the overall star formation
  efficiency is a {\em free parameter}. The physically motivated range
  roughly lies between 20\% to 60\%. For smaller values feedback
  processes are likely to be still too weak to significantly alter or
  halt star formation, and for larger values the collective effects of
  protostellar outflows, winds, and UV radiation from massive stars
  (in the case of massive clusters) will have modified the star
  forming cloud so dramatically that the simple gas laws adopted in
  most cluster forming calculations break down.\\
  When using coordinates and velocities of protostars from cluster
  forming calculation as input for the subsequent cluster evolution,
  one usually makes the further {\em assumption} of instantaneous gas
  removal, because the stellar dynamics calculations typically neglect
  any contributions of gas to the cluster potential.
\item {\em Close binaries:} Most cluster formation calculations can only
  describe the formation of very wide binaries which essentially form by a
  capturing process when two gas clumps each containing one protostar merge
  together. Close binaries may form from gravitational instabilities in
  protostellar accretion disks, which are not resolved in typical cloud
  fragmentation calculations (see, however, one high resolution calculation by
  Bate {\it et al.} 2002). If accretion disks are not resolved in a
  calculation, one needs to {\em assume} a close binary fraction and assign
  mass ratios and orbital eccentricities to each core in the simulation before
  forwarding this information to the cluster evolution code.
\item {\em Small $N$:} The current molecular cloud fragmentation
  calculations are at best able to describe the formation of clusters
  with about a hundred stars. This is insufficient for most stellar
  dynamics purposes. However, with further advances in computer
  technology and with improved parallel algorithms, modeling the
  formation of star clusters with thousands of members will become
  possible in the near future. One will be able to follow the 
  evolution of star clusters like the Trapezium Cluster in Orion, or
  of the Pleiades, or Hyades fully ab initio. For larger stellar
  clusters there is no hope to consistently include star formation into
  the stellar dynamic calculations. One needs to resort to 
  theoretical considerations as discussed in Section
  \ref{subsec:init-theoretical}. 
\item {\em Pre-main sequence evolution:} Very massive stars may enter
  the stellar main sequence while still accreting being deeply
  embedded in their parental gas cocoon.  Low-mass protostars, on the
  contrary, spend a long time in the classical pre-main sequence
  contraction phase (e.g.\ Palla \& Stahler 1999).  For the MODEST approach
  this means that even long after gas expulsion, during the first several
  millions of years  of star cluster evolution, stellar dynamics not only needs
  to be combined with stellar evolution modules for main sequence
  stars, but also pre-main sequence modeling needs to be included.
  This is the more important as pre-main sequence stars have
  considerably larger stellar radii than stars on the main sequence and
  therefore are more susceptible to collisional processes.
\end{enumerate}

\subsection{Initial Conditions from Theoretical Considerations}
\label{subsec:init-theoretical}

Whereas the initial conditions for star
clusters with a small number of members $N$ can be motivated by star
forming calculations as advocated before, this is not true for very
massive star clusters.  It is not possible to simply `scale up' the
properties of small-$N$ clusters into the large number regime.
Therefore, initial conditions for large-$N$ clusters mostly will be
obtained by searching for an appropriate distribution function $f$.

In the absence of any information from computer-generated models, we
may distinguish between theoretical equilibrium configurations and
more realistic cluster models based on observations of star-forming
regions.

At the simplest level, Plummer (1911) models are often used.  However, these
have no direct connection with dynamics and should therefore be
considered as convenient models for test purposes.

Theoreticians often like to investigate families of well-defined
models.  A wide variety of equilibrium models can be described using
King-Michie distributions.  Adopting a distribution function of the
type $f(E,J_z)$, we can generate a sequence of models both in terms of
the central concentration parameter and the amount of rotation.  In
addition, velocity anisotropy can also be considered (see, e.g., Binney \&
Tremaine 1987).

Initial subclustering forms a good alternative to idealized
distributions and may actually be more useful for realistic cluster
simulations.  The origin of runaway stars gives rise to one
interesting set of problems posed by this scenario.

Given the initial coordinates and velocities, we also need to specify
an initial mass function and there is a wide choice for the
latter.  For practical purposes, a piecewise fitting function based on
observational data may be adopted.  This still leaves the question of
the upper mass limit which plays an important role.  Initial mass
segregation presents a further uncertainty, although there are some
observational constraints.

Primordial binaries represent another important ingredient of star
cluster simulations.  The main parameters here can be summarized as
$f(a,e,m_1/m_2)$, where the semi-major axis ($a$) distribution needs
to span many decades.  The distribution of eccentricities ($e$) may be
of secondary importance but the mass ratios ($m_1/m_2$) have a direct
bearing on the end-point of binary evolution.  Likewise the upper mass
limit affects the production of degenerate objects which are known to
reside in clusters.  It is also worth emphasizing that cluster
evolution is speeded up significantly by the presence of a mass
spectrum.

Finally, a non-equilibrium value for the initial virial ratio leads to
violent relaxation and core-halo formation on a short time scale.  Such
models may be relevant in connection with removal of the remaining gas.

%
%
%
%

\section{Data Structures and Formats}
\label{datastructures}

\subsection{Exchanging Data}

Communication among the various independent
modules of a running program may be accomplished via simple functional
(subroutine) interfaces which define and strictly control how much one
module needs to know about the workings of another.  It is equally
desirable for separate programs to communicate in some standard way.
One can easily imagine situations where we wish to compare the
operations of, or simply share data between, two programs implementing
alternative treatments of the same underlying physics.  Examples might
be Monte-Carlo and $N$-body treatments of stellar dynamics, SPH and
recipe formulations of fluid mechanics, or different sets of recipes
for stellar and binary evolution.  For purposes of comparison and
communication, it is essential that these programs all be able to
interpret and manage the same input data sets.

Such a requirement immediately raises several significant technical
problems.  Simply put, different programs may generate very different
kinds of data, organized internally in unique, even conflicting, ways,
and possibly sampled inhomogeneously in space or time.  For example, a
2-D stellar evolution code might produce as output a series of
two-dimensional arrays representing various thermodynamic quantities
at uniformly spaced sampling times, or the sampling intervals may be
chosen so that the data tend to cluster around interesting
evolutionary stages.  A simple SPH simulation with shared time steps
would generate identical data for all particles, sampled uniformly in
time but non-uniformly in space.  An $N$-body simulation with variable
time steps naturally generates inhomogeneous data (different data for
different particles), sampled non-uniformly in both space and time,
often with some sort of hierarchical (tree) structure implicit in the
data.  In the grand simulations contemplated here, we must allow for
the possibility of any and all of these data formats (and others!)
being freely mixed in the I/O stream.

Some basic design considerations then are: (1) How do we accommodate a
broad range of data formats in a flexible way?  (2) Should we
distinguish between complete data streams used to reconstruct entire
calculations and much simpler ``snapshot'' files used to checkpoint
and restart simulations?  (3) How much data should be saved in a file,
and how much should be recomputed when the file is read?  (4) How do
we represent the data in space and in time?  The choice leads to the
interpretation of position and time as either array indices or
particle attributes.  We must be able to support both descriptions.
(5) How do we represent particle attributes at each sampling point,
however defined: as an array of physical quantities (homogeneous
particle data), or as a collection of tagged properties (inhomogeneous
data)?  (6) Finally, since we expect to be dealing with very large
amounts of data, how can we accomplish these goals efficiently?  We
discuss some examples of data formats (FITS, NEMO, Starlab/story,
Starlab/tdyn,...),\footnote{See \tt http://www.manybody.org/modest.html}
each of which addresses one or more some of the above points.

\subsection{The {\tt tdyn} Data Representation}

A simple ``snapshot'' data format is adequate for many types of
simulation.  Systems with small dynamic range are easily synchronized
(or are synchronous by construction), making it both straightforward
and efficient to save data at regular predefined intervals.  Not all
calculations lend themselves to this approach, however.  $N$-body
simulations, for example, naturally produce data in a quite different
format.  Their large dynamic range means that individual particle time
steps are the norm, meaning that particle trajectories are updated
non-uniformly, each at the ``right'' rate, as defined by its own local
time scales.  A complete description of the dynamical evolution
requires that we find a convenient way of saving and reproducing this
level of detail in a data file.

Of course, one can easily produce snapshots of an $N$-body system for
check-pointing and restart purposes.  In practice, time steps are
chosen to be powers of two, greatly improving scheduling efficiency by
organizing particles into blocks which can be updated simultaneously.
This also means that the system is necessarily synchronized at regular
intervals, allowing snapshots to be produced.  (Even without block
time steps, synchronization can be forced at any desired output time,
but block steps are widely used and facilitate the process.)  However,
for many purposes, such as analyzing particle motion or visualizing
the evolving data set, it is desirable to reproduce trajectories in
detail.  Snapshots are poorly suited to this task, as they are
severely limited by the dynamic range of the data, which generally
makes interpolation impossible unless the interval between snapshots
is made impracticably small.

We have developed a data structure that allows us to save and
manipulate enough detail to reproduce the native $N$-body structure
without significant loss of resolution.  In essence, rather than
trying to sample the positions and velocities of stars at fixed,
finely spaced intervals, we provide these data asynchronously, at key
points along each orbit, then use this information to interpolate each
trajectory to any specified time.  As an analogy, if we wished to
provide a means of specifying the locations of all trains within the
New York subway system at any given time, we probably would not opt to
publish a long list of all train positions on a second-by-second
basis.  Rather, it would be much more efficient to provide a timetable
stating when trains arrive and depart from each station, together with
some simple rule for computing a train's movement en route from one
station to the next.  The description below is couched in the
terminology of the Starlab environment within which this approach was
developed.  However, the basic ideas are common to all $N$-body codes.

In the simplest approach, we might save particle positions and
velocities at the end of every $N$-body step.  Then, to determine the
particle's position at any intermediate time, we use a fourth-order
interpolation scheme (in fact, the same scheme used in the $N$-body
code) to fit the position and velocity at each end of the stored
interval spanning the desired time.  In this way we can reconstruct a
continuous, differentiable trajectory that reproduces the original
$N$-body track.  The choice of time step ensures that the sampling
interval is adequate.  As a practical matter, we find that sampling
every step is unnecessary.  Sub-sampling the trajectory---saving data
only every 20--30 time steps---allows adequate reproduction for most
purposes, ensuring energy conservation to better than 0.1\%.  The
resulting volume of saved data is large (about 100 bytes per particle
per $N$-body time unit), but manageable given the proper tools, provided
with Starlab.

As the simulation proceeds, particle data are saved in a more or less
unstructured way, as follows.  At the end of each chosen step, the
system simply writes a self-contained record specifying particle ID,
mass, position, velocity, and other properties (e.g.~stellar
evolutionary state) to the output stream.  The task of reconstituting
this ``stream of consciousness'' into a usable data structure is left
entirely to the program reading and interpreting the data.  As a
practical matter, it is convenient to print complete snapshots of the
system at regularly spaced synchronization times, typically separated
by a few dynamical time scales.  The complete external representation
then consists of segments of data starting and ending with full
snapshots defining the hierarchical tree structure of the entire
$N$-body system, connected by asynchronous sequences of particle records
spanning each trajectory.  These data segments, typically a few tens
megabytes in size for 10k particles, form the basic unit of external
data.

Changes in tree structure resulting (for example) from binary
formation and destruction are handled in a manner analogous to the
segments forming the full data set.  When two particles combine to
form a binary, the event is signaled by terminating records marking
the end of the individual particle trajectories, followed by a
snapshot of the new binary tree (center of mass plus components)
marking the start of a new trajectory family.  The reverse occurs when
a binary splits back into components.  Thus all structural changes in
the tree, large and small, are clearly delineated by ``bookends''
defining the old and new tree structures.

Internally, as the stored data are read in, they are assembled into a
four-dimensional tree structure, mimicking the standard Starlab
linked-list describing spatial structure at any given time, but with
the added dimension of forward and backward pointers in time allowing
navigation along a given trajectory.  (The {\tt dyn} in {\tt tdyn}
refers to the basic Starlab data structure; the {\tt t} refers to
time.)  Determining a particle's position at any given time then
amounts to identifying its trajectory and the interval along it
spanning the desired time, then interpolating from the saved position
and velocity data to obtain the desired information.  By construction,
the tree structure at any instant is completely defined by the tree
structure on the most recent snapshot along each particle's
trajectory.

We note that most non-dynamical data vary slowly, or may even be
constant, across a given interval or segment, so simple linear
interpolation is usually adequate.  Finally, since step-by-step
treatment of internal binary motion generally requires too much
storage in this treatment, most binaries are treated as evolving
``kepler'' structures describing their slowly varying orbital
elements.  A binary is reconstituted by first interpolating its
orbital structure, then locating its components in the interpolated
orbit.

\subsection{Mixed Particle Type Snapshots}

Here we would like to argue that for data exchange
(initial conditions, archiving, Virtual Observatory, etc.), one
should use a simple, lowest common denominator
description of how the data will be stored ``offline''.
This means leaving out the
complexities of varying time in the dataset, and adopt the
notion of a snapshot.


\begin{figure}[t]
  \centerline{\epsfxsize=3.0in \epsfbox{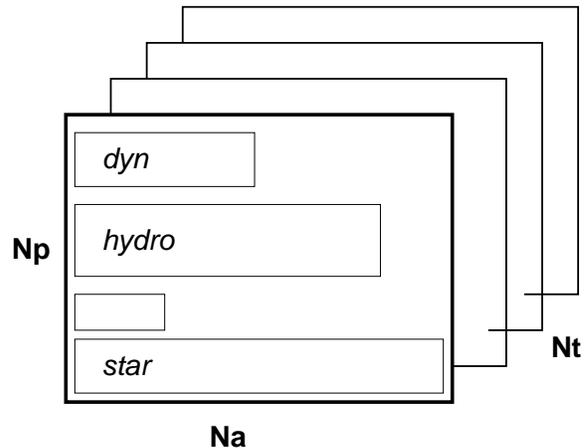}}
  \caption{A ``PAT'' cube, 
with for $Na$ attributes in $Np$ particles in $Nt$ SnapShots. 
Shown here are 4 SnapShots, with each 4 different type of Families
(dyn, hydro, no-name and star). Hierarchically:
Attribute $\in$ Particle $\in$ Family $\in$ SnapShot $\in$ Album}
\label{fig:patcube}
\end{figure}

Thinking of a snapshot as a matrix, where the columns are particle
attributes and the rows individual particles, a programmer
still has the choice to write the matrix column- or row major wise.
In addition, modern $N$-body codes often deploy a small number of 
different types of particles (e.g. pure gravity, SPH, stellar
evolution etc.) which can also evolve one type into another
and create new types as the system evolves.
So we arrive at a description
of a snapshot as a set of differently sized matrices, 
organized in the following hierarchy:

\footnotesize\begin{verbatim}
Attribute = a named Quantity (can also be vector)
Particle  = set of Attributes
Family    = set of Particles (same set of Attributes)
SnapShot  = set of Families
Album     = set of SnapShots
\end{verbatim}\normalsize

\noindent
where at each level a number of ``header'' variables are needed
to describe the items and the lower level items. 

At this stage we do not
want to suggest a particular implementation for
that data-format yet, 
as there are several possibilities, including possibly a new one.
Ideally we probably would want a self-descriptive
format, such as 
XML\footnote{See e.g. {\tt http://xml.gsfc.nasa.gov/XDF/XDF\_home.html}}.
NEMO's dataformat is also of this nature, and
one can equally well think of implementations in 
FITS (BINTABLE) and HDF, both of which
have had previous proposals floating around.
(Teuben 1995
\footnote{see also {\tt http://zeus.ncsa.uiuc.edu:8080/data\_format/}
{\tt data\_format.html}
for a proposal by Bryan \& Summers})

\subsubsection{Naming Conventions}

Apart from deciding the basic layout of the data, 
all header variables and columns need to get names
on which everybody can agree and give the same meaning to.
Instead of the usual Fortran unformatted I/O

\footnotesize\begin{verbatim}
WRITE (UNIT)  ((POS(K,J),K=1,3),J=1,NBODY)
\end{verbatim}\normalsize

we would envision some structured I/O routines, 
which could look as follows

\footnotesize\begin{verbatim}
CALL NBAWRITE (UNIT,'Position',POS,NBODY,3)
\end{verbatim}\normalsize

\subsubsection{Attribute (Column) Names}

As an initial suggestion here are some potential names one could
agree on that would give meaning to the associated data in
a snapshot:

\footnotesize\begin{verbatim}

// SnapShot and Family header variables

int    Npoint        //  9223372036854775808
int    Ndim          //  3     (number of dimensions)
int    Nattributes   //  3     (number of attributes)
string FamilyName    // 'disk', 'halo', 'gas', 'bulge'
string CoordSys      // 'cartesian-xyz', 'polar-rtp'
string CodeName      // 'arik'
string CodeAuthor    // 'Roald Teuben'
string CodeVersion   // '3.0'
string CodeDate      // '23-jun-2014'
string Hardware      // 'grape12'

// Particle Attributes (-vector means Ndim applies)

real-vector Pos[]           // simple dynamics
real-vector Vel[]
real-vector Pos[]
real        Mass[]
real-vector Acc[]
real        Potential[]
real        Density[]

real  SPHEntropy[]          // gas properties (SPH)
real  SPHTemperature[]      //
real  SPHSmoothingLength[]  // needs SPH kernel type
real  SPHDensity[]          // 
real  SPHAcc[]              // pressure gradient
int   SPHNneib[]            // number of neighbors

real  Age[]                 // stellar evolution
real  Temperature[]
real  Metallicity[]
real  Radius[]

real  SemiMajorAxis[]       // orbital elements
real  Eccentricity[]
real  Inclination[]
real  LongAscendNode[]
real  LongPeriapse[]
real  TrueLong[]
   
\end{verbatim}\normalsize

\section{Conclusions}
\label{conclusions}

\newcommand{\myskip}{\medskip}
\newcommand{\myem}{\myskip\noindent\em}

Dynamical simulations of dense star clusters have reached the point
where detailed treatments of many aspects of stellar physics must be
included.  A significant fraction of stars in globular clusters and
galactic nuclei are expected to experience close encounters or actual
physical collisions with other stars at some time during the evolution
of their parent system.  At the same time, collisions and the effects
of stellar and binary evolution can strongly influence cluster
dynamics, and may lead to the formation of objects whose properties
provide key insights into a cluster's past.  Population synthesis
studies have reached a similar conclusion from the opposite direction:
dynamical interactions can be vitally important in determining the
observed properties of dense stellar systems.

The dynamics of dense stellar systems is also essential for
understanding star cluster formation. While protostars in a dense
cluster environment build up, they are likely to interact strongly or
even merge, and in general they will compete with each other for gas
accretion. This has important consequences for the stellar mass
spectrum and for the subsequent dynamical evolution of the cluster.

In the workshop MODEST-1 (for MOdeling DEnse STellar systems) the
participants discussed many possible avenues for combining stellar
physics with stellar dynamics.  Options considered ranged from simple
rules and heuristic recipes, to extensive look-up tables using
precomputed data, to full-blown ``live'' simulations of stellar and
binary evolution and stellar hydrodynamics embedded in a dynamical
code.  The following is a consensus view of the current state of the
art and an assessment of feasible future developments in the various
subfields represented at the meeting.

{\myem Dynamics.} Traditionally, treatments of stellar and binary
evolution and simple recipes for collisions have been realized as
modules attached to existing dynamical integrators.  In part, this is
historical---dynamicists have had the most pressing reasons to
incorporate these effects into their simulations.  However, it is also
a fairly natural way to proceed, as the dynamical portion of a large
N-body calculation is usually also the part principally concerned with
large-scale structure, scheduling, and the orchestration of ``local''
events, such as binary formation and destruction, stellar
interactions, stellar evolution, and so on.  One might imagine
constructing a more democratic system in which the dynamics, stellar
evolution, and hydrodynamics are handled on an equal footing.  However
it seems likely that, for the foreseeable future, the dynamical
integrator will continue to provide the framework within which other
physical effects are incorporated.

{\myem Evolution of isolated stars.} For ``canonical'' stars that
start their lives on the main sequence with more or less normal
compositions and never experience close encounters with other stars,
there seems to be no strong reason to perform on-the-fly computations
of stellar evolution.  Such calculations will almost certainly be of
lower precision and contain less physics than existing published
calculations.  Rather, the most practical approach involves the use of
look-up tables and fitting formulae based on precomputed tracks,
essentially as already implemented in current N-body codes.

{\myem Evolution of isolated binaries.} Binary evolution is too
complex for live binary evolution programs, and is expected to remain
so for the foreseeable future.  No such programs currently exist, and
even simplified versions would likely be too fragile for standalone
use.  The physics can be very sensitive to small perturbations and in
many cases is not sufficiently well defined for encapsulation in a
program to be possible; the number of binary configurations in which
the detailed physics is simply unknown is depressingly large.  For the
same reasons, no definitive precomputed binary evolutionary tracks
exist.  The parameter space is probably too large for look-ups
analogous to those used in stellar evolution to be practical in any
case.  We thus expect continued use of recipes and heuristic rules of
increasing sophistication, again more or less as implemented in
existing N-body codes.  We note that this approach has the added
benefit of allowing an investigator to identify and parametrize key
binary properties, and to vary and study their effects in a controlled
way.

{\myem Hydrodynamics.} Some integrated treatment of stellar collisions
is clearly required.  Many collisions involving main-sequence stars
can be adequately handled by rules and recipes currently under
development, but it seems inevitable that others will have to be
performed on the fly, probably using SPH as the description of fluid
dynamics best suited to incorporation into a dynamical integrator.
Existing codes do not include such modules; most resort to
(over)simplified ``sticky'' criteria for stellar mergers.  Basic
self-contained SPH (or shortcuts such as entropy-sorting) treatments
of two-body collisions could in principle be added to existing codes
in a relatively straightforward way.  Integration of arbitrary stellar
encounters within a full N-body environment is probably a feasible,
but much longer-term, goal.

{\myem Collision Products.} Collisions---either direct, between unbound
stars, or indirect, resulting from binary evolution or temporary
capture of stars in binaries---will give rise to ``non-canonical''
stars quite unlike those normally studied by stellar evolution codes
or reported in the literature.  They will be out of thermal
equilibrium, will probably be rapidly rotating, and will have unusual
composition and entropy profiles.  We will not be able to precompute
and interpolate all the possibilities.  Here we really do need live
stellar evolution codes to study the appearance and evolution of the
collision products.  However, such studies pose a severe challenge to
existing techniques, and lie beyond the capabilities of current
stellar evolution codes.  The creation of a robust, standalone module
to handle the evolution of collision products is a high priority.

\myskip How to make the pieces communicate?  It is unrealistic to
expect researchers to completely rewrite their codes (no matter how
attractive such a prospect might be...) in order to merge them with
other programs.  Rather, it is better to create modular programs by
encapsulating parts or all of existing computer codes and define
robust interfaces specifying clearly the functionality of each module
and the data that must be provided and returned for each to work.
Such an approach is vital, as it will facilitate controlled comparison
of competing techniques.  Behind the interface, the structure of each
module will be entirely up to the programmer, so long as it conforms
rigorously to the agreed-upon interface specifications.  We have begun
a study of the interfaces, data structures, and communication
protocols needed to realize this goal.  The external representation of
simulation data is also an important and unresolved issue---we need to
share data between programs in an efficient, extendible, and
non-destructive way.

The first MODEST workshop was successful in bringing together the
three astrophysics communities of researchers working the fields of
stellar evolution, stellar hydrodynamics, and stellar dynamics.  We
will continue to hold these workshops twice yearly, thereby providing
a meeting point for those who are actively involved in simulating
dense stellar systems.  For further details, see the MODEST web
site.\footnote{\tt http://www.manybody.org/modest.html}


\bigskip

{\it Acknowledgments.}  We acknowledge the input of the participants
of the MODEST-1 workshop.  Here is the complete list of those who
attended part or all of the workshop:

\begin{table}[h]
\begin{tabular}{ll}
Sverre Aarseth     &  Jun Makino       \\
David Chernoff     &  Marc Hemsendorf \\
Scott Fleming      &  Rosemary Mardling \\
Marc Freitag       &  Steve McMillan \\
Yoko Funato        &  David Merritt \\
John Fregeau       &  John Ouellette \\
Mirek Giersz       &  Onno Pols \\
Paul Grabowski     &  Dina Prialnik  \\
Atakan G\"urkan    &  Fred Rasio \\
Jarrod Hurley      &  Helmut Schlattl \\
Piet Hut           &  Mike Shara \\
Vicky Kalogera     &  Shawn Slavin \\
Ralf Klessen       &  Rainer Spurzem \\
Attay Kovetz       &  Jerry Sussman \\
Yuexing Li         &  Peter Teuben \\
James Lombardi     &  Dany Vanbeveren \\
Mordecai Mac Low   &  Ron Webbink \\
\end{tabular}
\end{table}

In addition, we also acknowledge comments on the manuscript by Douglas
Heggie and Pavel Kroupa.
R.S.K. acknowledges financial support by the Emmy Noether
Program of the Deutsche Forschungsgemeinschaft (DFG, grant KL1358/1).
J.C.L. acknowledges support from NSF grant AST 00-71165.  His work was 
also partly supported by the National Computational Science Alliance under 
grant AST 98-0014N and utilized the NCSA SGI/Cray Origin2000 parallel 
supercomputer.
S.M. acknowledges support from NASA ATP grant NAG5-10775.
R.F.W.'s participation is supported in part by NASA grant
NAG 5-11016 to the University of Illinois.

\def\araa{{\em Ann.\ Rev.\ Astron.\ Astrophys.}}
\def\aas{{\em Astron.\ Astrophys.\ Suppl.\ Ser.}}
\def\aj{{\em Astron.\ J.}}
\def\anap{{\em Ann.\ Astrophys.}}   
\def\apj{{\em Astrophys.\ J.}}
\def\apjs{{\em Astrophys.\ J.\ Suppl.\ Ser.}}
\def\aap{{\em Astron.\ Astrophys.}}
\def\jcam{{\em J.\ Comput.\ Appl.\ Math.}}
\def\jcp{{\em J.\ Comput.\ Phys.}}
\def\jfm{{\em J.\ Fluid Mech.}}
\def\mnras{{\em Mon.\ Not.\ R.\ Astron.\ Soc.}}
\def\nat{{\em Nature}}
\def\pta{{\em Phil.\ Trans.\ A.}}
\def\ptp{{\em Prog.\ Theo.\ Phys.}}
\def\prd{{\em Phys.\ Rev.\ D}}
\def\prl{{\em Phys.\ Rev.\ Lett.}}
\def\prsa{{\em Proc.\ R.\ Soc.\ London A}}
\def\pasp{{\em Pub.\ Astron.\ Soc.\ Pac.}}
\def\zp{{\em Z.\ Phys.}}
\def\za{{\em Z.\ Astrophys.}}

\end{document}